\documentclass[twocolumn]{autart} %\documentclass[journal,twoside,web]{ieeecolor}
\usepackage{hyperref}
\usepackage{cite}
\usepackage{mathptmx} % assumes new font selection scheme installed
\usepackage{amsmath}
\usepackage{amssymb,amsfonts}
\usepackage{algorithmic}
\usepackage{graphicx}
\usepackage{textcomp}
\usepackage{xcolor}
\usepackage[scr]{rsfso}
\usepackage{soul}
 
\begin{document}
\begin{frontmatter}

\title{Neural Operators of Backstepping Controller and Observer\\ Gain Functions for Reaction-Diffusion PDEs}
\author%[author1]
{Miroslav Krstic}\ead{krstic@ucsd.edu},\ \ \
\author%[author2]
{Luke Bhan}\ead{lbhan@ucsd.edu},  \ \ \
\author%[author3]
{Yuanyuan Shi}\ead{yyshi@eng.ucsd.edu}  
\address%[author1,author2,author3]
{University of California, San Diego, USA}
%\vspace*{-3em}
        %\thanks{The authors are with the University of California, San Diego, USA, {krstic@ucsd.edu}, lbhan@ucsd.edu, yyshi@eng.ucsd.edu }

%\maketitle

%\vspace*{-2em}

\begin{abstract}
Unlike ODEs, whose models involve system matrices and whose controllers involve vector or matrix gains, PDE models involve functions in those roles---functional coefficients, dependent on the spatial variables, and gain functions dependent on space as well. The designs of gains for controllers and observers for PDEs, such as PDE backstepping, are mappings of system model functions into gain functions. These infinite-dimensional nonlinear operators are given in an implicit form through PDEs, in spatial variables, which need to be solved to determine the gain function for each new functional coefficient of the PDE. The need for solving such PDEs can be eliminated by learning and approximating the said design mapping in the form of a neural operator. Learning the neural operator requires a sufficient number of prior solutions for the design PDEs, offline, as well as the training of the operator. In recent work, we developed the neural operators for PDE backstepping designs for first-order hyperbolic PDEs. Here we extend this framework to the more complex class of parabolic PDEs. The key theoretical question is whether the controllers are still stabilizing, and whether the observers are still convergent, if they employ the approximate functional gains generated by the neural operator. We provide affirmative answers to these questions, namely, we prove stability in closed loop under gains produced by neural operators. We illustrate the theoretical results with numerical tests and publish our code on \href{https://github.com/lukebhan/NeuralOperatorsForAdvectionDiffusionControl}{github}. The neural operators are three orders of magnitude faster in generating gain functions than PDE solvers for such gain functions. This opens up the opportunity for the use of this neural operator methodology in adaptive control and in gain scheduling control for nonlinear PDEs. 
\end{abstract}

% \begin{IEEEkeywords}\normalfont
% PDE backstepping, machine learning, DeepONet
% \end{IEEEkeywords}

\date{}
\end{frontmatter}
\allowdisplaybreaks

%\vspace*{-1em}

\section{Introduction}

\paragraph*{ML as a tool for learning control methodologies.}
In the recent manuscript \cite{https://doi.org/10.48550/arxiv.2302.14265} we introduced a  learning-based control {\em framework} which devises a new role for machine learning (ML): learn an entire control design {\em methodology}, in the form of a mapping from the plant model to the controller gains, or even to the control inputs. 

This framework is neither model-free nor methodology-agnostic. On the contrary, it is method-specific. For a particular method (LQR, pole placement, MPC, backstepping, etc.), after a large number of training calculations of the controller gains on a sample set of plant models, an ML-approximated mapping of the method is learned. Once learned as a plant-to-gains mapping, the control design for a {\em new}/next plant (outside of the training set) does not require another solution of the design equations (Riccati, Bezout, etc.) but merely entails an ``evaluation of the learnt map'' to obtain the control gains. 

One would argue that no dire need exists for LQR or other linear finite-dimensional designs for such a learning-based capability, where an entire methodology is ``encoded'' into a neural mapping. The cost of the solution of a design problem (say, a Riccati equation, even of a high dimension) is not prohibitive, even online with current technology. Indeed, we are not motivated by design challenges in finite dimensions but for PDEs. 

In PDE control, the design problems are not matrix equations. They are PDEs themselves (or harder problems, such as operator Riccati equations). Since the infinite-dimensional state of a PDE is a function of {\em spatial} variables, the controller gain is also a function of spatial variables. Finding the gain typically entails solving a PDE in space (but not in time). 
%While solving a matrix equation may no longer be a numerically formidable problem, solving a PDE for a new plant may well be a numerical problem that is complex enough that it stands in the way of an application of the PDE control methodology. 
It is therefore of interest, in PDE control, to have a capability where producing the control gain functions is just an evaluation of a neural mapping that has already learned the design methodology on a large set of previously offline-solved control design problems for a sample set of PDEs in a certain class. 

\vspace*{-1em}\paragraph*{Neural operators for approximating mappings of functions into functions.} Just as %it is appropriate to understand 
the control designs for linear finite-dimensional systems are matrix-to-matrix mappings ($A, B$ into gain $K$), 
%of matrices (such as the plant data $A$ and $B$) into matrices (such as the control gain $K$), 
control designs for PDEs %should be understood as 
are function-to-function mappings (spatially-dependent coefficients into gains). 
Our inspiration for encoding PDE control methodologies into machine learning comes from recent  advances in the mathematics of machine learning. Motivated by the tasks of finding solution/flow maps (from the initial conditions into future states) for physical PDEs (such as the difficult Navier-Stokes PDEs), research teams led by George Karniadakis~\cite{lu2019deeponet,lu2021deeponet},  Anima Anandkumar and Andrew Stuart~\cite{li2020neural, li2021fourier}, and George Pappas and Manfred Morari~\cite{kissas2022loca,seidman2022nomad}, have developed neural approximation methods,  termed ``neural operators,'' with provable properties for nonlinear operators acting on functions and producing functions. These approaches are not simply discretizing PDEs and finding solution maps to the resulting large ODE solution problems. In the language of distributed parameter systems, they are not ``early lumping'' methods of learning solution maps. They approximate (non-discretized) function-to-function nonlinear operators and provide guarantees of the accuracy of approximation in terms of the required sizes of the training sets and neural networks. 

The value of such a capability in PDE control cannot be understated. With a theoretically rigorous and numerically powerful capability like this, specific PDE control methods, for specific classes of PDEs, can be learned once and encoded as neural operators, ready to produce the control gain functions for any new functional coefficients of the same classes of PDEs. 

In a theoretically rigorous field like PDE control, a computational capability with rigorous approximation guarantees has a value primarily if it allows the retention of the theoretical properties proven for the ``exact design''. This is indeed what we show in the paper \cite{https://doi.org/10.48550/arxiv.2302.14265}  in which we introduce the framework: approximate neural operator representations of a particular PDE control method---PDE backstepping---preserves its stability guarantees in spite of the control gains not being generated by solving the design PDEs but by the gains being generated from the learned ``neural model'' of PDE backstepping.

\begin{figure}[t]
    \centering \includegraphics[width=2.6in]{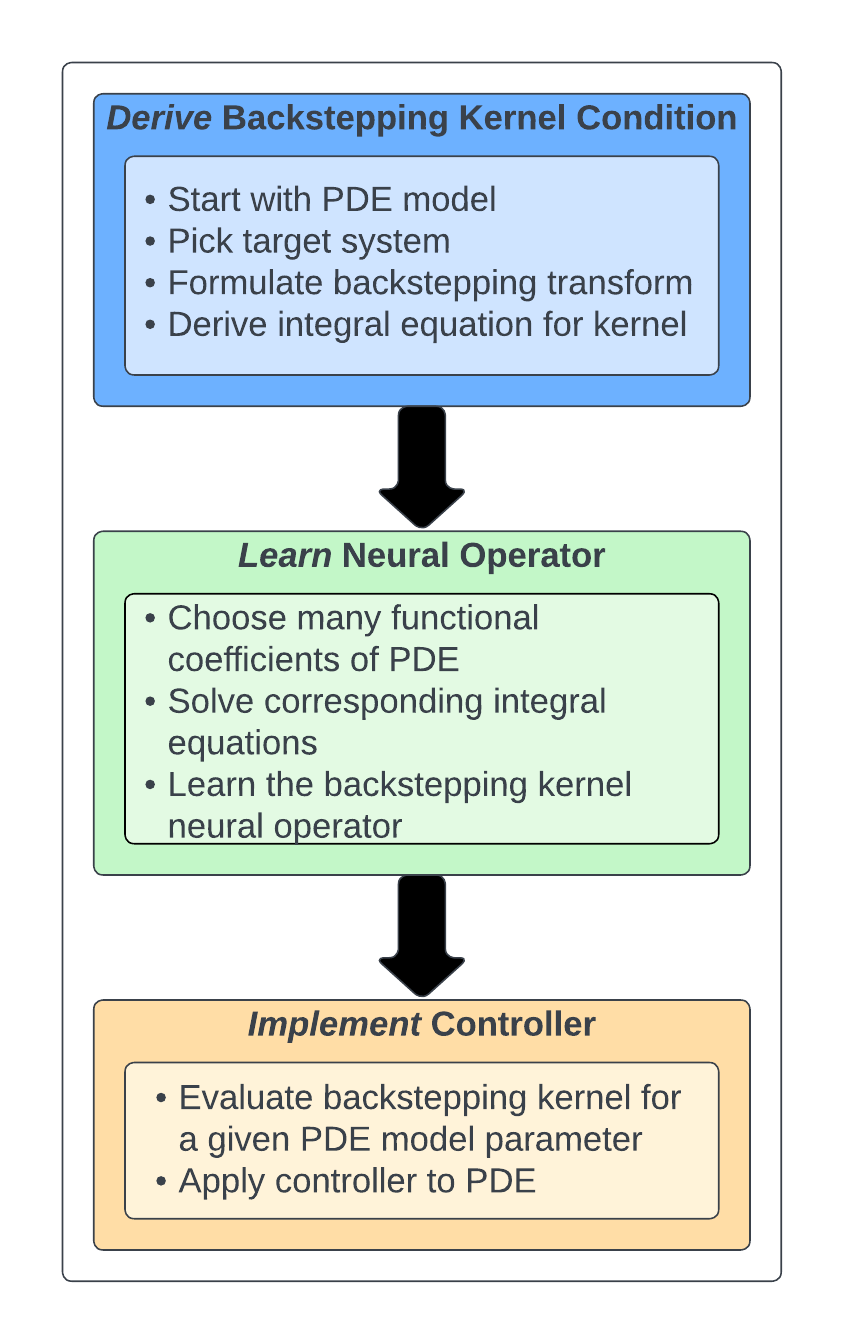}
	\caption{An algorithmic representation of our design paradigm of employing neural operators in boundary control of PDEs. Three major step clusters are performed: (1)  \underline{derivation} of the integral equations for the backstepping kernels, performed only once; (2)  \underline{learning} of the mapping from the plant parameter functions into the backstepping kernel functions, also performed only once; and (3) \underline{implementation} of the controller for specific plant parameters. The task in the top box has  been completed in \cite{krstic2008Backstepping,1369395}. In this paper, the task in the middle box is introduced and stability guarantees for the task in the bottom box are provided.}
    \label{fig:0}
\end{figure}

\begin{figure} 
 %   \centering
\hspace*{-2em}    \includegraphics[scale=0.9]
{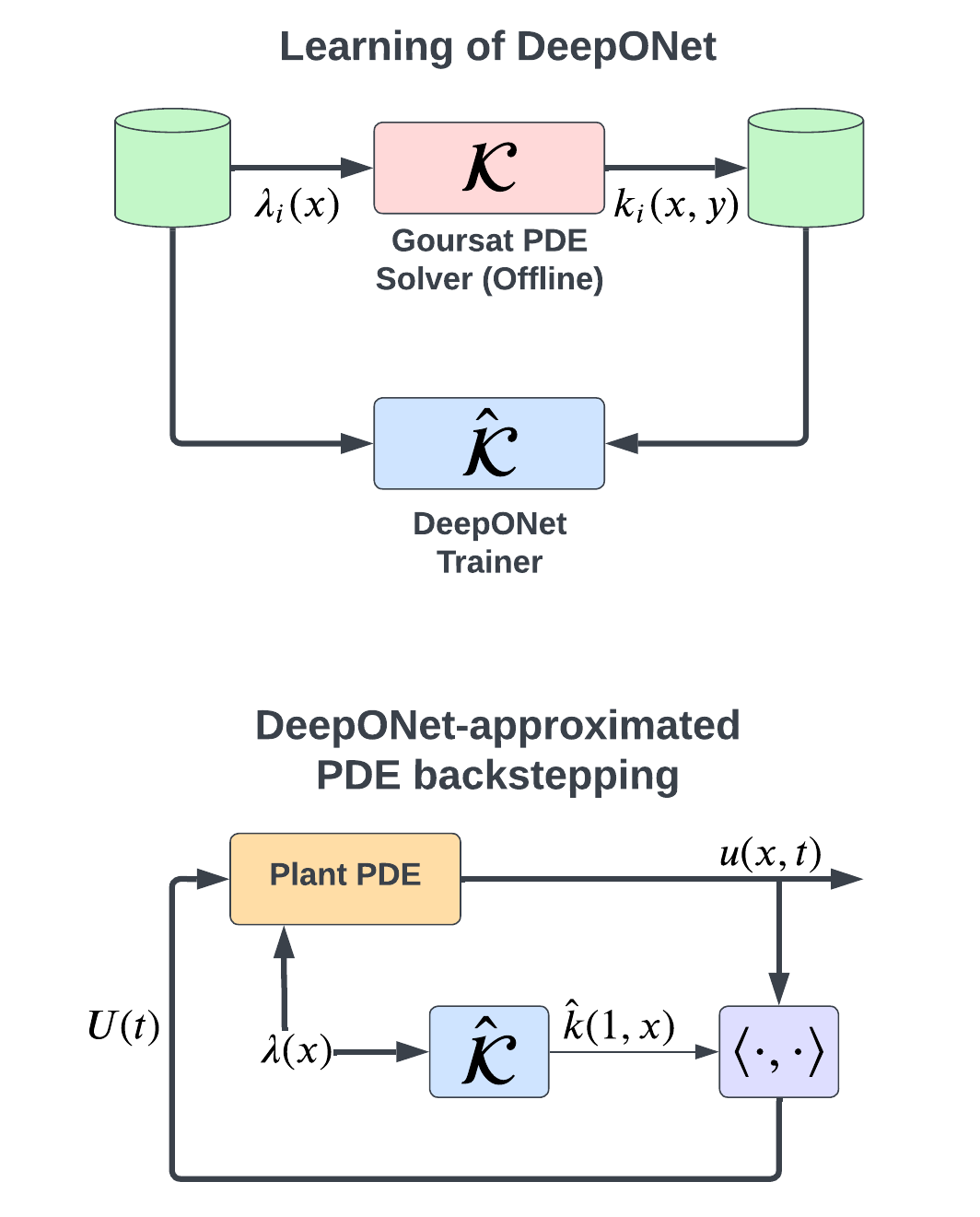}
    \caption{Stages (2) and (3) of the framework in Figure \ref{fig:0}. TOP: The process of learning the PDE backstepping design operator  $\mathcal{K}:\lambda \mapsto k$ involves many solutions of a kernel PDE $k_{xx}-k_{yy} = \lambda k$ in the Goursat form, for different functions $\lambda_i(x)$ and then training of a neural operator $\hat{\mathcal{K}}:\lambda\mapsto \hat k$. BOTTOM: Feedback implementation of PDE backstepping control with gain kernel $\hat k(1,x)$ generated by the DeepONet $\hat{\mathcal{K}}$.}
    \label{fig:deeponetdiagram}
\end{figure}

\vspace*{-1em}\paragraph*{Extension of PDE backstepping neural operators from hyperbolic \cite{https://doi.org/10.48550/arxiv.2302.14265}  to parabolic PDEs.} Hyperbolic PDEs involve only the first derivatives in space and time. This makes them (all else being equal) the ``simplest'' PDE class for control. Delay systems combine ODEs with delays---the simplest form of a PDE. While the simplest among PDEs, hyperbolic PDEs are not necessarily easy to control. They can be unstable, with many unstable eigenvalues, and only one input acting at the boundary of a domain. This mix of simplicity within the PDE family, with the non-triviality for control, makes hyperbolic PDEs the ideal entry point for any new study in PDE control, including the introduction of a new framework for learning-based control in our \cite{https://doi.org/10.48550/arxiv.2302.14265}. The framework is depicted in Figure \ref{fig:0}. The {\em learning} and {\em implementation} portions of the framework in Figure \ref{fig:0} are depicted in Figure \ref{fig:deeponetdiagram}.

Control design problems for hyperbolic PDEs are hyperbolic PDEs themselves, namely, equations with only first derivatives in multiple spatial variables. Parabolic PDEs, with their first derivative in time but second derivatives in space, are the natural next challenge for learning the PDE backstepping methodology using neural operators. This is what we undertake in this paper. The chief difficulty with learning backstepping kernel operators for parabolic PDEs is that the kernels are governed by second-order PDEs, which raises the difficulty for solving such PDEs and for proving the sufficient smoothness of their solutions so that the neural operator (NO) approximations have guarantee of sufficient accuracy for preserving stabilization. At the intuitive level, with more derivatives, and more boundary conditions, the nonlinear operator from the reaction function to the gain in parabolic PDEs is a more complex operator than the nonlinear operator from the recirculation function to the gain in hyperbolic PDEs. There are hyperbolic cases where the kernel mapping can be written (though not solved) using the Laplace transform. That is never the case with parabolic PDEs; the design problem is never of spatial dimension lower than two. 

\begin{figure*}
    \centering
    \includegraphics{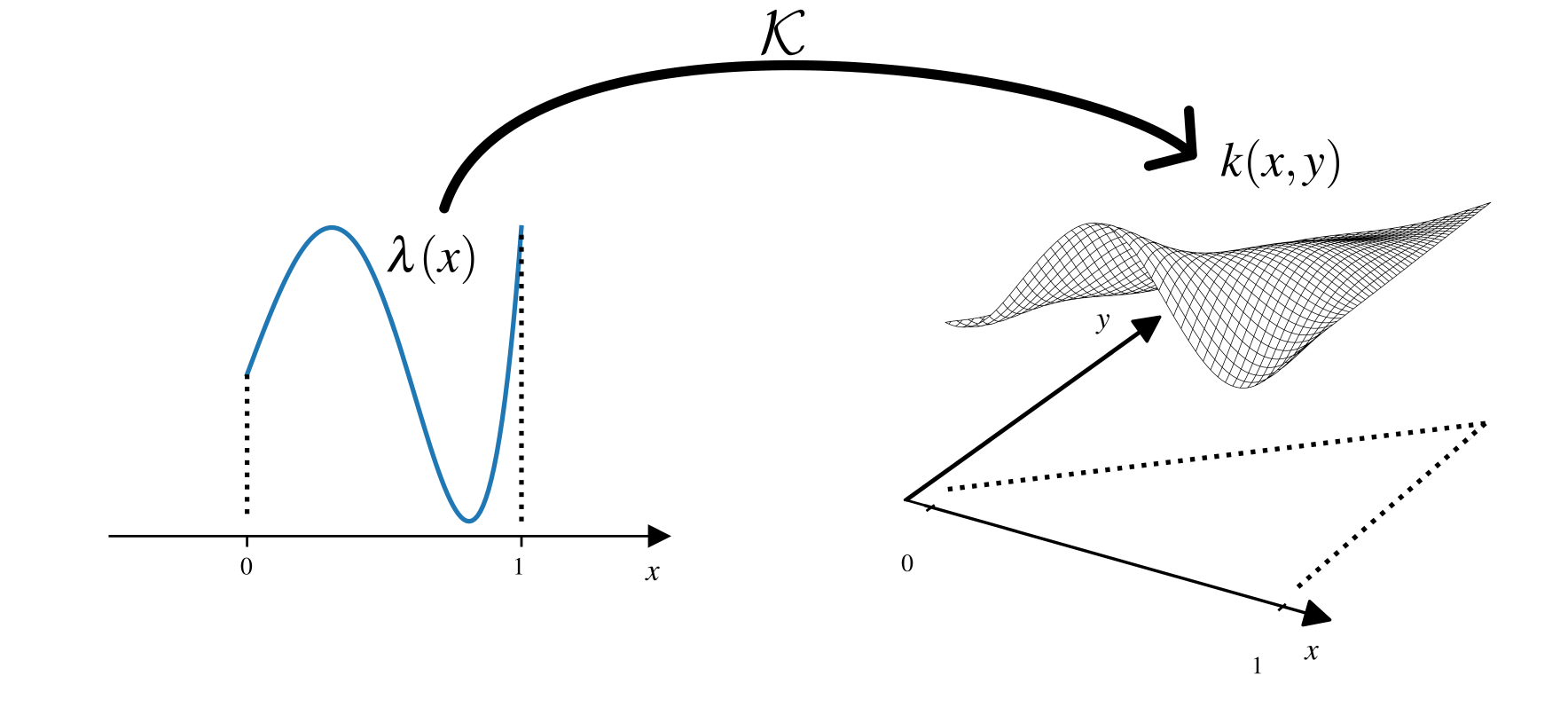}
    \caption{The PDE backstepping design operator ${\mathcal K}: \lambda \mapsto k$, where $\lambda(x)$ is the spatially-varying reaction coefficient of the PDE, whereas $k(x,y)$ is the kernel function of the backstepping transformation, producing the feedback gain function $k(1,y)$ in the feedback law $U(t) = \int_0^1 k(1,y) u(y,t) dy$.}
    \label{fig:my_label}
\end{figure*}

We consider parabolic PDE systems of the form
\begin{eqnarray}
    \label{eq-PDE}
	u_t(x,t) &=& u_{xx}(x,t) + \lambda(x) u(x, t) , \qquad \mbox{$x\in[0,1)$}
 \\  \label{eq-PDEBC0}
 u(0,t) &=& 0
 \\  \label{eq-PDEBC1}
	u(1, t) &=& U(t). 
\end{eqnarray}
Our goal is the design of a PDE backstepping boundary control 
%law of the integral (in space) form
\begin{equation}\label{eq-bkstfbkk}
    U(t) = \int_0^1 k(1,y) u(y,t) dy,
\end{equation}
as well as an observer with the (collocated) boundary sensing of $u_x(1,t)$. By ``design'' we mean to find the gain function $k$ in the control law \eqref{eq-bkstfbkk}, namely, to find the output $k$ of the function-to-function mapping $\mathcal{K}:\lambda \mapsto k$, depicted in Figure \ref{fig:my_label}. This paper's objective is to learn the design operator $\mathcal{K}$ with a neural operator approximation $\hat{\mathcal{K}}$ (top of Figure \ref{fig:deeponetdiagram}) and to employ the resulting approximate gain $\hat k$ in the control law (bottom of Figure \ref{fig:deeponetdiagram}).

\begin{table}[t]	
\centering
\begin{tabular}{|c|c|c|l}
% \cline{2-5}
% & \multicolumn{2}{c|}{deterministic}
% & \multicolumn{2}{c|}{stochastic} 
\cline{1-3}
actuation & opposite boundary & sensing
%\\ \cline{1-5} 
\\ \hline \hline
$u(1,t)=U(t)$ & $u(0,t)=0$ & $u_x(0,t)$
& anti-col
\\ \cline{1-3}
$\pmb{u(1,t)=U(t)}$ & $\pmb{u(0,t)=0}$ & $\pmb{u_x(1,t)}$
& {\bf col}
\\ \cline{1-3}
$u(1,t)=U(t)$ & $u_x(0,t)=0$ & $u(0,t)$
& anti-col
\\ \cline{1-3}
$u(1,t)=U(t)$ & $u_x(0,t)=0$ & $u(1,t)$
& col
\\ \cline{1-3}
$u_x(1,t)=U(t)$ & $u(0,t)=0$ & $u_x(0,t)$
& anti-col
\\ \cline{1-3}
$u_x(1,t)=U(t)$ & $u(0,t)=0$ & $u_x(1,t)$
& col
\\ \cline{1-3}
$u_x(1,t)=U(t)$ & $u_x(0,t)=0$ & $u(0,t)$
& anti-col
\\ \cline{1-3}
$u_x(1,t)=U(t)$ & $u_x(0,t)=0$ & $u(1,t)$
& col
\\ \cline{1-3}
\end{tabular}
\caption{Eight possible combinations of boundary actuation, sensing, and boundary condition at the opposite end of $[0,1]$. We focus on the simplest combination---in the second row.} 
\label{table-contributions}
\end{table}

Since parabolic PDEs in one dimension have two boundary conditions, and also boundary actuation and boundary sensing can be employed at either boundary, a total of sixteen combinations of boundary actuation, boundary sensing, and boundary condition on the unactuated boundary are possible. Taking the symmetry between the boundaries $x=0$ and $x=1$ into account, the total number of truly distinct combinations is eight. They are listed in Table \ref{table-contributions}. 

We are able to solve all eight problems but, in this paper, pursue the simplest of the eight combinations for pedagogical reasons. The case with Dirichlet boundary conditions, $u(0,t)=0, u(1,t)=U(t)$ is, notationally, the simplest case. It allows the reader to most quickly grasp the utility and the technical steps in employing neural operators in the control of parabolic PDEs.  

All the results in the paper---a full-state controller, an observer, and an output-feedback law (as well as the seven additional combinations not pursued in the paper)---can be extended to the more general class of parabolic PDE systems,
\begin{eqnarray}
  v_t(x,t) &=& \varepsilon(x)v_{xx}(x,t) + b(x) v_x(x,t) 
  \nonumber\\
  &&+ \lambda(x) v(x,t) + g(x) v(0,t) 
  \nonumber\\
  && +\int_0^x f(x,y) v(y,t) dy, \quad x\in(0,L). 
\end{eqnarray}
 The domain $[0,L]$ is easily normalized to $[0,1]$, the diffusion $\varepsilon(x)$ is easily normalized to unity, and the advection term $b(x) v_x(x,t)$ is easily eliminated with a scaling transformation. 
%All we perform here with a Dirichlet boundary condition at $x=0$ can be done with Neumann or Robin boundary conditions. Likewise, control at $x=1$ can be converted from Dirichlet to Neumann. 
We forego pursuing this myriad of alternatives for pedagogical reasons---they are overwhelming but standard. Likewise, we forgo the treatment of the terms $g(x) v(0,t) +\int_0^x f(x,y) v(y,t) dy$ because it adds complications but it is standard as well.

The parabolic PDE with unity diffusion and with spatially-varying reaction $\lambda(x)$ is a perfect introduction to the possibilities that neural operators present for PDE backstepping control where the computation of the gain kernel $k(x,y)$ for    each new $\lambda(x)$ can be avoided by developing a neural operator approximation of the functional mapping (nonlinear operator) ${\mathcal K}: \lambda \mapsto k$, which is depicted in Figure \ref{fig:my_label}. 

We opt to present in the paper the results for the combination in the second row of Table \ref{table-contributions} because this combination allows us to ``kill two birds with one stone'' in our exposition. For this particular actuator-sensor combination, which is collocated (and the simplest of the four collocated combinations), the same kernel is used to obtain the gain functions for both the controller and the observer. This relieves  the reader of the burden of following multiple approximations of the kernel, multiple neural operators, multiple training processes for those operators, and multiple theorems that guarantee the approximability of those multiple operators. The concept of encoding the methodologies of controller, observer, and output-feedback design into a neural operator is grasped through a single operator ${\mathcal K}: \lambda \mapsto k$. And the duplication in the exposition is avoided. A reader who possesses the skills in calculations and the stamina can work out the remaining combinations in Table \ref{table-contributions}.

\vspace*{-1em}\paragraph*{PDE Backstepping.}
Even though PDE backstepping was first developed for parabolic systems \cite{1369395}, it is best to begin its study from the easier, hyperbolic case \cite{krstic2008Backstepping}. Control of hyperbolic PDEs has grown into a rich area because, in the hyperbolic case, one can stabilize a coupled system with fewer inputs than PDEs. A {\em pair} of coupled hyperbolic PDEs was stabilized with a single boundary input in \cite{Coron2013Local}, an  extension to $n+1$ hyperbolic PDEs with a single input was introduced in \cite{Meglio2013Stabilization},  an extension to $n+m$  PDEs with boundary actuation on $m$ ``homodirectional'' PDEs in \cite{Hu2016Control,hu2019boundary},  an extension to cascades with ODEs  in \cite{DIMEGLIO2018281}, and an extension to ``sandwiched'' ODE-PDE-ODE systems in \cite{WANG2020109131,9319184}. Redesigns robust to delays are provided in \cite{Auriol2018Delay1}. PDE backstepping-based output-feedback regulation with disturbances is proposed in \cite{DEUTSCHER201556,deutscher2018}.

For parabolic PDEs, backstepping for full-state feedback stabilization was developed in \cite{1369395} and for observer design in \cite{smyshlyaev2005backstepping}. A complex extension from linear to nonlinear parabolic PDEs, using infinite Volterra series, was provided in \cite{vazquez2008control1,vazquez2008control2}. Backstepping was combined with differential flatness in \cite{MEURER20091182}. The first solutions to the null-controllability problem for parabolic PDEs were provided, using backstepping, in \cite{coron2017null,ESPITIA2019398}. Sampled-data and event-triggered versions of backstepping for parabolic PDEs appeared in \cite{karafyllis2018sampled,karafyllis2021event,espitia2021event,rathnayake2021observer}. Work on cascades of parabolic PDEs with other systems has included heat-ODE cascades \cite{krstic2009compensating,bekiaris2010compensating}, delay-parabolic cascades \cite{krstic2009control}, and ODE-heat-ODE sandwich systems \cite{wang2019output}. A backstepping design for a moving-boundary  PDE-ODE Stefan system was presented in \cite{s2016output}.
Coupled parabolic PDEs introduce special challenges and have been tackled in \cite{baccoli2015boundary,orlov2017output,vazquez2016boundary}. Extensions from multiple 1D parabolic PDEs to PDEs in 2D and higher dimensions, such as in the book \cite{meurer2012control} are arguably even more challenging and have been pursued for Navier-Stokes and magnetohydrodynamic systems in \cite{vazquez2007closed,vazquez2008magnetohydrodynamic} on channel domains, as well as for reaction-diffusion systems on balls of arbitrary dimensions \cite{vazquez2016explicit}. 
Adaptive control designs for parabolic PDEs were introduced in \cite{krstic2008adaptive,smyshlyaev2007adaptive2,smyshlyaev2007adaptive3}, extended in \cite{karafyllis2019adaptive1}, and extended to the hyperbolic case in \cite{Bernard2014}.
For coupled hyperbolic PDEs with unknown parameters, the parabolic designs in \cite{krstic2008adaptive,smyshlyaev2007adaptive2,smyshlyaev2007adaptive3} inspired a comprehensive collection of adaptive control designs in the book \cite{Anfinsen2019Adaptive}. Applications of backstepping to PDE models of traffic are introduced in \cite{Yu2019Traffic,Yu2022}.

\vspace*{-1em}\paragraph*{Advances in learning-based control.}
What we present here is one more among many directions in learning-based control. For the benefit of the reader from PDE control, we highlight a few results from this vast and growing literature. Stability of learning-based MPC was established in \cite{ASWANI20131216,rosolia2017learning} and followed, for nonlinear systems, by efforts on joint learning of the controller and(or) Lyapunov functions \cite{tedrakeNNControl, chang2019neural, chen2020learning, chen2021learning, dawson2022safe}. 
In addition, \cite{taylor2019control, nguyen2021}
have explored how learning-based control affects systems with known Lyapunov functions,  
\cite{boffi2021learning, pfrommer2022tasil, claudioNonlinear2023} studied learning of stability certificates and stable controllers  from data,
%\cite{nguyen2021, taylor2019control}. 
% Furthermore, \cite{richards2018lyapunov, boffi2021regret} has explored learning adaptive stability certification with deep neural networks. 
and \cite{frank2022} developed a provably stable data-driven algorithm based on system measurements and prior system knowledge. 

 For reinforcement learning (RL) \cite{Bert05}, the focus has been on learning the system dynamics and 
 providing closed-loop guarantees in \emph{finite-time} for both linear  \cite{dean2018regret,chen2021black,lale2022reinforcement} and nonlinear systems~\cite{berkenkamp2017safe,kakade2020information,singh2021learning,lale2022kcrl}. For model-free RL, \cite{fazel2018global,mohammadi2021convergence,jiang2022,zhao2023global} proved the convergence of policy optimization to the optimal controller for LTI systems, \cite{PANG2020109035, pramod2022} for LTV  systems, ~\cite{tang2021analysis} for partially observed linear systems. For a review of policy optimization (PO) methods for LQR, $H_{\infty}$ control, risk-sensitive control, LQG, and output feedback synthesis, see \cite{hu2022towards}. For nonlinear systems, \cite{chow2018lyapunov, cui2022structured,shi2022stability} investigated PO with stability guarantees from CLFs. In addition to PO, \cite{vamvoudakis2010online, lewis2012reinforcement, bhasin2013novel, vamvoudakis2017q} proved stability and convergence of actor-critic methods~\cite{vamvoudakis2010online, lewis2012reinforcement} and Q-learning~\cite{vamvoudakis2017q}.
In CPS, learning-based control was developed for partially observable systems \cite{andreasSeparation2023}.

Learning-based control in games and for MAS is pursued in \cite{zhang2019policy, mazumdar2020gradient, fiez2020implicit, zhang2020model,mao2022provably, poveda2022fixed, vamvoudakis2017game, qu2020scalable}. Convergence is shown to Nash equilibria in zero-sum linear quadratic games~\cite{zhang2019policy}, continuous games~\cite{mazumdar2020gradient}, Stackelberg games~\cite{fiez2020implicit}, Markov games~\cite{zhang2020model,mao2022provably}, and multi-agent learning over networked systems~\cite{qu2020scalable, poveda2022fixed}. 

We focus on learning-based control for PDE systems. In our previous work~\cite{shi2022machine}, we demonstrate the empirical success of using NOs for accelerating PDE backstepping observers. Our recent work~\cite{https://doi.org/10.48550/arxiv.2302.14265} represents the first step towards using NOs for \emph{provably} bypassing gain computations and directly learning the controller with closed-loop stabilization guarantee, in hyperbolic PDE systems. 
%A  review for learning-based control in games is in~\cite{zhang2021multi}.

\vspace*{-1em}\paragraph*{Neural operators---a brief summary.}
Neural operators are neural network-parameterized maps for learning relationships between function spaces. They consist of three components: an encoder, an approximator, and a reconstructor \cite{lanthaler2022errorDeeponet}. The encoder is an interpolation from an infinite-dimensional function space to a finite-dimensional vector representation. %For example, in our work we consider the pointwise evaluations of our function on the interval $[0, 1]$. 
The approximator aims to mimic the infinite map using a finite-dimensional representation of both the domain function space and the target function space. The reconstructor then transforms the approximation output into the infinite-dimensional target function space. 
The implementation of both the approximator and the reconstructor is generally coupled and can take many forms. For example, the original DeepONet \cite{lu2021deeponet} contains a ``branch'' net that represents the approximation network and a ``trunk'' net that builds a basis for the target function space. The outputs of the two networks are then taken in linear combination with each other to form the operator. FNO \cite{li2021fourier} utilizes the approximation network in the Fourier domain where the reconstruction is done on a basis of the trigonometric polynomials. LOCA \cite{kissas2022loca} integrates the approximation network and reconstruction step with a unified attention mechanism. NOMAD \cite{seidman2022nomad} extends the linear reconstructor map in DeepONet to a nonlinear map that is capable of learning on nonlinear submanifolds in function spaces.

With the basic notions and notation for NOs given in Appendix \ref{sec-NN}, we state next the key technical result that enables our use of NOs to learn the PDE backstepping kernel mappings. The result is quoted in its general/abstract form. It is specialized to the PDE control setting in our Theorem \ref{cor-DeepONet}. 

\begin{thm}\label{thm-DeepONet}
{\em (DeepONet universal approximation theorem \cite[Theorem 2.1]{lu2021advectionDeepONet}).}
Let 
$X \subset \mathbb{R}^{d_x}$ and 
%$K_1 \subset \mathbb{R}^{d_1}$ 
 $Y \subset \mathbb{R}^{d_y}$ be compact sets of vectors $x\in X$ and $y\in Y$, respectively. 
Let ${\mathcal U}:X\rightarrow U\subset \mathbb{R}^{d_u}$ and ${\mathcal V}:Y\rightarrow V\subset \mathbb{R}^{d_v}$ be sets of continuous functions $u(x)$ and $v(y)$, respectively. 
Let ${\mathcal U}$ be also compact.
% \subset C^0(X; U)$
% %$V$ 
% be a compact subset of functions $u(x)$ %in $C^0(X; U)$, 
% with values in $U\subset \mathbb{R}^{d_u}$.  
% %$C^0(K_1)$, 
% %$Y \subset \mathbb{R}^d$ 
% %$K_2 \subset \mathbb{R}^d$ 
% %be a compact set 
% Let  
% ${\mathcal V}=p C^0(Y;V)$ be the set of all functions $v(y)$ continuous in $y\in Y$, with values in $V\subset \mathbb{R}^{d_v}$. 
%$Y=C^0(K_2)$. 
Assume  
the operator $\mathcal{G}: {\mathcal U} \rightarrow {\mathcal V}$ 
%$\mathcal{G}: V \rightarrow Y$ 
is %H\"{o}lder 
continuous.
Then, for all $\epsilon > 0$, there exist $m^*, p^* \in \mathbb{N}$ such that for each $m \geq m^*$, $p \geq p^*$, there exist $\theta^{(k)}, \vartheta^{(k)}$, neural networks $f^{\mathcal{N}}(\cdot ; \theta^{(k)}) , g^\mathcal{N}(\cdot ; \vartheta^{(k)}),  k=1,\ldots, p$, and $ x_j \in X,  j=1, \ldots, m$, with corresponding $\mathbf{u}_m = (u(x_1), u(x_2), \cdots, u(x_m))^{\rm T}$, such that
\begin{equation} \label{eq:2.14G}
|\mathcal{G}(u)(y) - \mathcal{G}_\mathbb{N}(\mathbf{u}_m)(y)| < \epsilon %\hspace{10pt} 
%\nonumber\\ &&
%\mathcal{G}_\mathbb{N}(\mathbf{u}_m) = \sum_{k=1}^p g^\mathcal{N} (\mathbf{u}_m ; \Theta^{(k)}) f^\mathcal{N} (y; \theta^{(k)})
\end{equation}
for all functions $u \in {\mathcal U}$ and all values $y \in Y$ of ${\mathcal G}(u)\in{\mathcal V}$. 
\end{thm}

In the sequel, we denote the DeepONet neural operator values $\mathcal{G}_\mathbb{N}(\mathbf{u}_m)(y)$ compactly as $\hat{\mathcal G}(u)(y)$ and the operators themselves as $\hat{\mathcal G}$.

\vspace*{-1em}\paragraph*{Paper outline and contributions.}
In Section \ref{sec-bkst-intro} we recap the basic PDE backstepping approach from \cite{1369395}. Recalling in Section  \ref{sec-kernelLip} the twice continuous differentiability of the backstepping kernel function
%in Section \ref{sec-bkst-kernel}. In Section \ref{sec-kernelLip}, 
we establish the existence of a neural operator with an arbitrary accuracy for a set of continuously differentiable reaction coefficients not exceeding a certain size in the supremum norm. Sections \ref{sec-bkst-full-state} and \ref{sec-bkst-observer} contain our main results. In Section \ref{sec-bkst-full-state} we prove the stability of a feedback law employing a DeepONet approximation of the backstepping gain. In Section \ref{sec-bkst-observer} we prove the convergence of a backstepping observer that employs a DeepONet approximation of the observer gain. In Section \ref{sec-bkst-output-feedback} we combine the DeepONet-based full-state feedback and observer, to obtain a DeepONet-based output feedback controller with an actuator-sensor pair collocated at the $x=1$ boundary. In Section \ref{sec-numerical} we illustrate the theoretical results with numerical tests. 

This paper's contribution relative to the inaugural work on backstepping for parabolic PDEs \cite{1369395} is in providing a methodology for capturing the backstepping design in the form of a neural operator and avoiding the need for the solution of kernel PDEs, after the neural operator is once synthesized. This capability is highly valuable in future work in the adaptive control of parabolic PDEs and gain-scheduling for semilinear parabolic PDEs. In relation to our recent work on neural operator approximated backstepping control of hyperbolic PDEs, this paper extends this methodology, including stability guarantees, to a more difficult class of PDE systems and kernel operators. Additionally, compared to \cite{https://doi.org/10.48550/arxiv.2302.14265} where only full-state feedback is considered, in this paper, we solve problems in observer design and output-feedback control, with a convergence guarantee for the DeepONet-approximated backstepping observer.

\section{Basic Backstepping Design for Reaction-Diffusion PDE}
\label{sec-bkst-intro}
We employ the following backstepping transformation, 
\begin{eqnarray}
 	w(x, t) = u(x, t) - \int_0^x k(x,y)u(y, t) dy,
\end{eqnarray}
to convert \eqref{eq-PDE}, \eqref{eq-PDEBC0}, \eqref{eq-PDEBC1}
into the target system 
\begin{eqnarray}\label{eq-targetPDE}
	 w_t &= &w_{xx} \\ \label{eq-targetPDEBC0}
	w(0,t) &=& 0\\ \label{eq-targetPDEBC1}
	w(1,t) &=& 0
\end{eqnarray}
with the help of feedback
\eqref{eq-bkstfbkk}. We could as well pursue the target system $w_t =w_{xx}-cw, c>0$, but we forego this design flexibility for the sake of simplicity.

To convert \eqref{eq-PDE}, \eqref{eq-PDEBC0}, \eqref{eq-PDEBC1} into \eqref{eq-targetPDE}, \eqref{eq-targetPDEBC0}, \eqref{eq-targetPDEBC1}, $k$ needs to satisfy
\begin{eqnarray}
\label{eq-kPDE}
k_{xx}(x,y) - k_{yy}(x,y) &=& \lambda(y) k(x,y), \quad \forall (x,y)\in \breve{\mathcal T}
\\ \label{eq-kPDEBC0}
k(x,0) &=& 0
\\ \label{eq-kPDEBCx}
k(x,x) &=& -{1\over 2} \int_0^x \lambda(y) dy
\end{eqnarray}
where $\breve{\mathcal T} = \{0< y\leq x <1\}$ and ${\mathcal T} = \{0\leq y\leq x \leq 1\}$. 

The following assumption is important.

\begin{assum}\label{ass-lambda}
$\lambda \in C^1([0,1])$. 
\end{assum}

% \section{Properties of the Backstepping Kernel}
% \label{sec-bkst-kernel}

\section{Accuracy of Approximation of Backstepping Kernel Operator with DeepONet}
\label{sec-kernelLip}

%We denote ${\mathcal K}: \lambda \mapsto k$, which is a mapping from $C^1[0,1]$ to $C^2({\mathcal T}) $. 

\begin{thm}{\em (proven in \cite{1369395,Smyshlyaev2010})}
\label{thm-kC2}
For every $\lambda \in C^1([0,1])$, the PDE system \eqref{eq-kPDE}, \eqref{eq-kPDEBC0}, \eqref{eq-kPDEBCx} has a unique $C^2({\mathcal T})$ solution with the property
\begin{equation}\label{eq-kbound}
|k(x,y)|\leq \bar\lambda {\rm e}^{2\bar\lambda x},
\end{equation}
%$k(x,0)=0$, and $\left|{d\over dx}(k(x,x))\right| \leq \bar\lambda/2$ for any $\lambda \in C^1[0,1]$, 
for all $x\in[0,1]$, where $\bar\lambda = \sup_{x\in [0,1]}|\lambda(x)|$. 
% In particular, $\partial_2 k \in C^0({\mathcal T})$, where 
% \begin{equation}
% \partial_2 := \partial_{xx} - \partial_{yy}, 
% \end{equation}
% and $|k_{xx}(x,y)- k_{yy}(x,y)|\leq \bar\lambda^2 {\rm e}^{2\bar\lambda x}$.
\end{thm}

This theorem is proven by representing the PDE system \eqref{eq-kPDE}, \eqref{eq-kPDEBC0}, \eqref{eq-kPDEBCx}  as an integral equation
\begin{eqnarray}\label{eq-Ginteq}
G(\xi,\eta) &=& -{1\over 4} \int_\eta^\xi \lambda\left({s\over 2}\right) ds
\nonumber \\
&& + {1\over 4} \int_\eta^\xi \int_0^\eta \lambda\left({\sigma - s\over 2}\right)
G(\sigma,s) ds d\sigma, 
\end{eqnarray}
where 
\begin{eqnarray}
\label{eq-xyxieta}
\xi = x+y,\ \ \eta = x-y, && x ={\xi+\eta\over 2}, \ \ y = {\xi - \eta\over 2}
\ \ 
\\
G(\xi,\eta) = k(x,y) &=& k\left( {\xi+\eta\over 2},  {\xi - \eta\over 2}\right).
\end{eqnarray}
The change of variables \eqref{eq-xyxieta} converts the domain ${\mathcal T}$ for $(x,y)$ into the larger triangular domain ${\mathcal T}_1 = \{0\leq \eta\leq \xi\leq 1\} \cup \{ 1\leq \xi\leq 2-\eta \leq 2\}$ for $(\xi,\eta)$. The integral equation \eqref{eq-Ginteq} is one of the useful approaches in generating solutions for $k(x,y)$ for the purpose of training the neural approximation of the operator $\lambda\mapsto k$.

Next, denote the set of functions 
\begin{equation}
    \underline{K} = \left\{\left. k\in C^2({\mathcal T}) \right| k(x,0) = 0, \forall x\in[0,1]\right\}
\end{equation} 
and let the operator ${\mathcal K}:C^1[0,1]\rightarrow %C^2({\mathcal T})
\underline{K}$ be defined by
\begin{equation}
k(x,y) =: {\mathcal K}(\lambda)(x,y). 
%k =: {\mathcal K}(\lambda),
\end{equation}
% the operator ${\mathcal K}_2:C^1[0,1]\rightarrow C^0({\mathcal T})$ denote
% \begin{equation}
% \partial_2 k = {\mathcal K}_2(\lambda),
% \end{equation}
Additionally, let the operator ${\mathcal M}:C^1[0,1]\rightarrow %C^2({\mathcal T})
\underline{K} \times C^1[0,1] \times 
C^0({\mathcal T})
$ be defined by
\begin{equation}
\left(k(x,y),
%\lambda(x) + 2 {d\over dx}\left(k(x,x)\right) ,k_{xx}(x,y) - k_{yy}(x,y) - \lambda(y)k(x,y)
\kappa_1(x), \kappa_2(x,y) \right) =: {\mathcal M}(\lambda)(x,y),
%(k,\partial_2 k) = {\mathcal M}(\lambda). 
\end{equation}
where
\begin{eqnarray}
    \kappa_1(x) &=& 2 {d\over dx}\left(k(x,x)\right) +\lambda(x) 
    \\
    \kappa_2(x,y) &=& k_{xx}(x,y) - k_{yy}(x,y) - \lambda(y)k(x,y).
\end{eqnarray}
Based on Theorem \ref{thm-kC2}, $\mathcal M$ is a continuous operator. By applying Theorem \ref{thm-DeepONet}, we get the following key result for the approximation of a backstepping kernel by a DeepONet (top of Figure \ref{fig:deeponetdiagram}).

\begin{thm}
%{\em (to Theorem~\ref{thm-DeepONet}).}
\label{cor-DeepONet}
For all $B_\lambda , B_{\lambda'}>0$ and $\epsilon > 0$, there exists a neural operator $\hat{\mathcal M}$ such that, for all $(x,y)\in {\mathcal T}$, 
%$p^*(B,\epsilon), m^*(B,\epsilon) \in \mathbb{N}$, with an increasing dependence on $B_\lambda $ and $1/\epsilon$, such that for each $p \geq p^*$ and $m \geq m^*$ there exist $\theta^{(k)}, \vartheta^{(k)}$, neural networks $f^{\mathcal{N}}(\cdot ; \theta^{(k)}) , g^\mathcal{N}(\cdot ; \vartheta^{(k)}),  k=1,\ldots, p$, and $ x_j \in [0,1], j=1, \ldots, m$, with corresponding $\mathbf{\beta}_m = (\beta(x_1), \beta(x_2), \cdots, \beta(x_m))^{\rm T}$,
%%$f^{\mathcal{N}}(\cdot ; \theta_k)$ and $g^\mathcal{N}(\cdot ; \vartheta_k), \: x_j \in [0, 1], \: k=1,\ldots, p, \: j=1, \ldots, m$, 
%such that
\begin{equation} \label{eq-Meps}
\left|{\mathcal M}(\lambda )(x,y) - \hat{\mathcal M}(\lambda )(x,y)\right | < \epsilon \end{equation}
holds for all Lipschitz $\lambda$ with the properties that $\|\lambda\|_\infty\leq B_\lambda, \|\lambda'\|_\infty\leq B_{\lambda'}$, namely, there exists a neural operator $\hat{\mathcal K}$ such that $\hat{\mathcal K}(\lambda)(x,0) \equiv 0$ and 
\begin{align} \label{eq-Keps}
&\left|{\mathcal K}(\lambda )(x,y) - \hat{\mathcal K}(\lambda )(x,y)\right | 
\nonumber\\
&+ \left|2{d\over dx}\left({\mathcal K}(\lambda )(x,x) - \hat{\mathcal K}(\lambda )(x,x)\right)
\right| \nonumber\\
&+ \left|\left(\partial_{xx} - \partial_{yy}\right)\left({\mathcal K}(\lambda )(x,y) - \hat{\mathcal K}(\lambda )(x,y)\right)
\right. \nonumber\\
&\left.-\lambda(y)\left({\mathcal K}(\lambda )(x,y) - \hat{\mathcal K}(\lambda )(x,y)\right) \right| 
< \epsilon.
\end{align}
\end{thm}

% \section{Perturbed Target Systems}
% \label{sec-perttarget}

\section{Stabilization under DeepONet Gain Feedback}
 \label{sec-bkst-full-state}

The following theorem establishes the properties of the feedback system at the bottom of Figure \ref{fig:deeponetdiagram}.

\begin{thm}
\label{thm-stabDeepONet-gf}
Let $B_\lambda, B_{\lambda'}>0$ be arbitrarily large and consider the system  \eqref{eq-PDE}, \eqref{eq-PDEBC0}, \eqref{eq-PDEBC1} with any $\lambda\in C^1([0,1])$ whose derivative $\lambda'$ is Lipschitz and which satisfies
$\|\lambda\|_\infty\leq B_\lambda$ and $ \|\lambda'\|_\infty\leq B_{\lambda'}$.  
 There exists a  sufficiently small $\epsilon^*(B_\lambda, B_{\lambda'})>0$ such that the feedback law 
 \begin{equation}\label{eq-fbkgf}
    U(t) = \int_0^1 \hat k(1,y) u(y,t) dy,
\end{equation}
 with all NO gain kernels $\hat k = \hat{\mathcal K}(\lambda)$ of 
 %arbitrary desired 
   approximation accuracy $\epsilon\in (0,\epsilon^*)$ in relation to the exact backstepping kernel $k={\mathcal K}(\lambda)$ ensures that 
 %there exist $M, c^*>0$ such that 
 the closed-loop system satisfies the exponential stability bound 
 %\eqref{eq-expstabbound}.
%Thus we arrive at
\begin{equation}\label{eq-expstabbound}
\|u(t)\| \leq M {\rm e}^{-(t-t_0)/2} \|u_0\|, \quad\forall t\geq t_0,
\end{equation}
where
\begin{equation}\label{eq-Mepslam}
M(\epsilon,\bar\lambda)= 
\left(1+\bar\lambda {\rm e}^{2\bar\lambda}\right)
\left(1+\bar\lambda {\rm e}^{2\bar\lambda} +\epsilon\right)
 {\rm e}^{\bar\lambda {\rm e}^{2\bar\lambda} +\epsilon}.
\end{equation}
 \end{thm}

\begin{pf}
{\em Approximate backstepping transform and perturbed target system.}
Take the backstepping transformation 
\begin{equation}
    \hat w(x,t) = u(x,t) - \int_0^x \hat k(x,y) u(y,t) dy, 
\end{equation}
where $\hat k = \hat{\mathcal K}(\lambda)$. 
With the control law
\eqref{eq-fbkgf}, 
the target system becomes
\begin{eqnarray}
\label{eq-target-gf}
    \hat w_t(x,t) &=& \hat w_{xx}(x,t) 
    +\delta_{k0}(x) u(x,t)
    \nonumber\\
    && + \int_0^x \delta_{k1} (x,y) u(y,t) dy
    \\
    \hat w(0,t) &=&0
       \\
    \hat w(1,t) &=&0,
\end{eqnarray}
with
\begin{eqnarray}\label{eq-deltakxy0}
\delta_{k0}(x) &=& 2{d\over dx}\left(\hat k(x,x)\right) +\lambda(x)
\nonumber \\
&=& -2{d\over dx}\left(\tilde k(x,x)\right) 
\\
\label{eq-deltakxy1}
    \delta_{k1}(x,y) &=& \partial_{xx} \hat k(x,y)-\partial_{yy} \hat k(x,y) - \lambda(y) \hat k(x,y)
    \nonumber\\
    &=& - \partial_{xx} \tilde k(x,y)+\partial_{yy} \tilde k(x,y) + \lambda(y) \tilde k(x,y),
\end{eqnarray}
where 
\begin{equation}\label{eq-ktilde}
    \tilde k = k-\hat k = {\mathcal K}(\lambda) - \hat {\mathcal K}(\lambda).
\end{equation}
With \eqref{eq-Keps}, we get
\begin{eqnarray}
\label{eq-bounddeltak0}
    \|\delta_{k0}\|_\infty &\leq & \epsilon
    \\
\label{eq-bounddeltak1}
    \|\delta_{k1}\|_\infty &\leq & %(1+\bar \lambda) 
    \epsilon.
\end{eqnarray}
{\em Inverse approximate backstepping transformation.} Since the state $u$ appears under the integral in the $\hat w$-system \eqref{eq-target-gf}, in the Lyapunov analysis we need the inverse backstepping transformation 
\begin{equation}\label{eq-invbkstgf}
    u(x,t) = \hat w(x,t) + \int_0^x \hat l(x,y) \hat w(y,t) dy.
\end{equation}
It is shown in \cite{krstic2008boundary} that the direct and inverse backstepping kernels satisfy in general the relationship
\begin{equation}
\hat l(x,y) = \hat k(x,y) + \int_y^x \hat k(x,\xi) \hat l(\xi,y) dy.     
\end{equation}
The inverse kernel satisfies the following conservative bound
\begin{equation}
\label{eq-lhatboundk}
    \|\hat l\|_\infty \leq \|\hat k\|_\infty {\rm e}^{\|\hat k\|_\infty}.
\end{equation}
Since $\|k-\hat k\|_\infty <\epsilon$, we have that $\|\hat k\|_\infty \leq \|k\|_\infty+\epsilon$. With \eqref{eq-kbound} we get 
\begin{eqnarray}
\label{eq-khatbound}
   \|\hat k\|_\infty &\leq& \bar k +\epsilon
   \\
\label{eq-khatboundbar}   
   \bar k(\bar \lambda) &:= &\bar\lambda {\rm e}^{2\bar\lambda},  
\end{eqnarray}
and hence 
\begin{equation}\label{eq-lhatbound}
    \|\hat l\|_\infty \leq \left(\bar\lambda {\rm e}^{2\bar\lambda} +\epsilon\right) {\rm e}^{\bar\lambda {\rm e}^{2\bar\lambda} +\epsilon}.
\end{equation}
{\em Lyapunov analysis.} The Lyapunov functional
\begin{equation}
V = {1\over 2} \|\hat w\|^2
\end{equation}
has a derivative 
\begin{equation}\label{eq-Vdot}
\dot V = - \|\hat w_x\|^2 + \Delta_0+\Delta_1,
\end{equation}
where
\begin{eqnarray}
\Delta_0(t) &=& \int_0^1\hat w(x,t) \delta_{k0}(x) u(x,t) dx
\\
\Delta_1(t) &=& \int_0^1\hat w(x,t) \int_0^x \delta_{k1}(x,y) u(y,t) dy dx. 
\end{eqnarray}
With several straightforward majorizations, we get
\begin{eqnarray}\label{eq-Deltabound0}
\Delta_0 
&\leq & \|\delta_{k0}\|_\infty\left(1+\|\hat l\|_\infty\right) \|\hat w\|^2
\nonumber\\
&=& \|\delta_{k0}\|_\infty\left(1+\|\hat l\|_\infty\right) 2V. 
\end{eqnarray}
and 
\begin{eqnarray}\label{eq-Deltabound1}
\Delta_1 &= & \int_0^1\hat w(x)\int_0^y \hat w(y) \int_y^x \delta_k(x,\sigma)\hat l(\sigma,y) d\sigma dy dx
\nonumber\\
&& + \int_0^1 \hat w(x) \int_0^x \delta(x,y) \hat w(y) dy dx
\nonumber\\
&\leq & \|\delta_{k1}\|_\infty\left(1+\|\hat l\|_\infty\right) \|\hat w\|^2
\nonumber\\
&=& \|\delta_{k1}\|_\infty\left(1+\|\hat l\|_\infty\right) 2V. 
\end{eqnarray}
From \eqref{eq-Vdot}, \eqref{eq-Deltabound0}, \eqref{eq-Deltabound1}, \eqref{eq-lhatbound}, and Poincare's inequality, we get
\begin{equation}
\dot V \leq -{1\over 2} (1 - \delta^*) V,
\end{equation}
where
\begin{equation}
\label{eq-delta*}
\delta^*(\epsilon,\bar\lambda) = 2\epsilon
%(1+\bar \lambda)  
\left(1+\bar\lambda {\rm e}^{2\bar\lambda} +\epsilon\right) {\rm e}^{\bar\lambda {\rm e}^{2\bar\lambda} +\epsilon}
\end{equation}
%is obtained from \eqref{eq-bounddeltak}, \eqref{eq-lhatbound} and 
is an increasing function of $\epsilon, \bar\lambda$, with the property that $\delta^*(0,\bar\lambda) = 0$. 
Hence, there exists $\epsilon^*(\bar\lambda)$ such that, for all $\epsilon \in [0,\epsilon^*]$,
\begin{equation}
\dot V \leq -{1\over 4} V,
\end{equation}
namely, $V(t)\leq V_0 {\rm e}^{-(t-t_0)/4}$. 
From the direct and inverse backstepping transformations it follows that
\begin{equation}\label{eq-Vsandwich}
    {1\over 1+\|\hat l\|_\infty }\|u\| \leq \sqrt{2V} \leq \left(1+\|\hat k\|_\infty\right)  \|u\|. 
\end{equation}
In conclusion,
\begin{equation}
\|u(t)\| \leq \left(1+\|\hat l\|_\infty \right) \left(1+\|\hat k\|_\infty\right) {\rm e}^{-(t-t_0)/2} \|u_0\|.
\end{equation}
With \eqref{eq-khatbound}, \eqref{eq-khatboundbar}, \eqref{eq-lhatbound}, the proof is completed. \qed
\end{pf}

\section{Observer Design}
\label{sec-bkst-observer}

State estimators (observers) with boundary measurements can be formulated with four measurement choices on the interval $[0,1]$: the measured quantities can be $u(0,t), u_x(0,t), u(1,t), u_x(1,t)$. That leads to many possible problem formulations. The possibilities multiply once we note that, on the opposite boundary from the one at which measurement is conducted, one can have either a Dirichlet or Neumann (or even Robin) boundary condition. Our objective in this paper is not to solve all the possible problems. We are concerned only with illustrating how NOs can be combined with PDE observers. Hence, our choice among the many possibilities is the simplest choice, of the highest pedagogical value. 

Since our goals with observers are twofold---to estimate the unmeasured state but also to use it in output-feedback control for stabilization---our choice of measurement needs to be consistent with the actuation choice we have already pursued in this note, namely, Dirichlet actuation of $u(1,t)=U(t)$. So, we cannot use $u(1,t)$ for measurement but we can use $u(0,t), u_x(0,t), u_x(1,t)$.
We make the last among these three choices. We 
let the output $u_x(1,t)$ be measured. 

\begin{figure} 
 %   \centering
\hspace*{-0.5em}    \includegraphics[scale=0.83]
{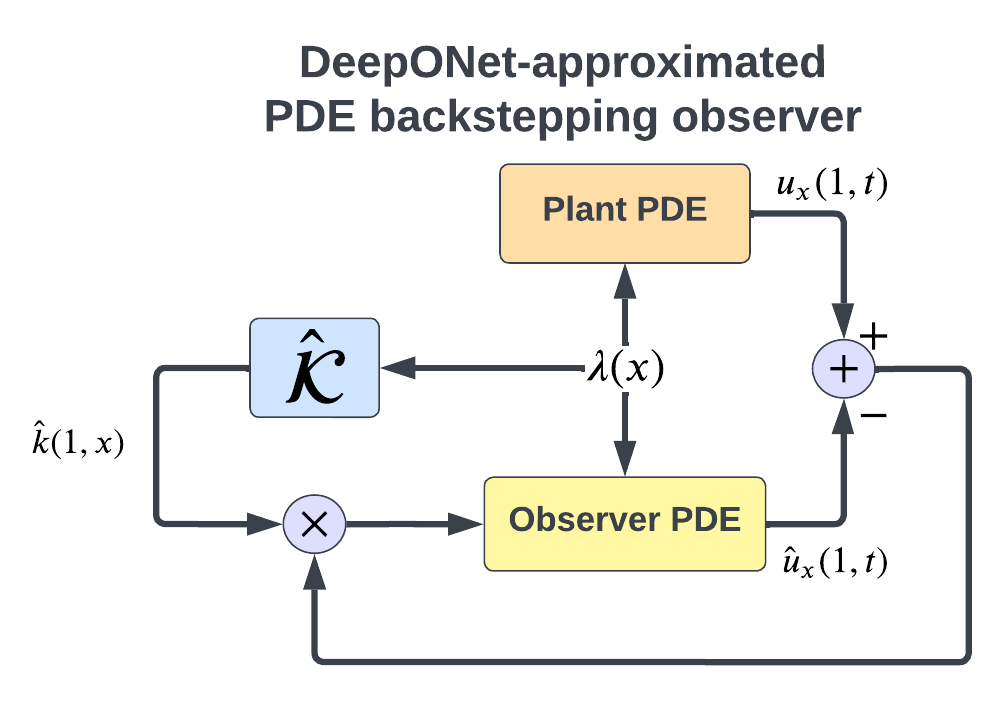}
    \caption{The PDE backstepping observer \eqref{eq-obsPDEkhat}, \eqref{eq-obsPDEBC0khat}, \eqref{eq-obsPDEBC1khat}  uses boundary measurement of the flux $u_x(1,t)$. The gain $\hat k(1,t)$ is produced with the DeepONet $\hat{\mathcal{K}}$.}
    \label{fig:observerdiagram}
\end{figure}

Our choice of $u_x(1,t)$ for measurement, as indicated in the observer diagram in Figure \ref{fig:observerdiagram}, is motivated by the fact that, with this measurement, an observer can be built using the same kernel $k(x,y)$ as for the control law. In other words, with a single training of a neural operator $\hat{\mathcal K}$, we obtain gains for both a controller and an observer---we kill two birds (pedagogically speaking) with one stone. We don't have to expend an undue amount of the reader's effort on the verification of the conditions of the DeepONet approximation theorem. It is enough for the reader to see once how this is done. The rest of the effort is better spent on illustrating the uses of this approximation capability in observers and output-feedback controllers.

\begin{thm}
\label{thm-stabDeepONet-gf-obs}
Let $B_\lambda, B_{\lambda'}>0$ be arbitrarily large and consider the system  \eqref{eq-PDE}, \eqref{eq-PDEBC0}, \eqref{eq-PDEBC1} with any $\lambda\in C^1([0,1])$ whose derivative $\lambda'$ is Lipschitz and which satisfies
$\|\lambda\|_\infty\leq B_\lambda$ and $ \|\lambda'\|_\infty\leq B_{\lambda'}$.  
 There exists a  sufficiently small $\epsilon^*(B_\lambda, B_{\lambda'})>0$ such that the observer
 \begin{eqnarray}
    \label{eq-obsPDEkhat}
	\hat u_t(x,t) &=& \hat u_{xx}(x,t) + \lambda(x) \hat u(x, t) 
 \nonumber\\
 &&- \hat k(1, x)\left[u_x(1,t) - \hat u_x(1,t)\right]
 %, \qquad \mbox{$x\in[0,1)$}
 \\  \label{eq-obsPDEBC0khat}
 \hat u(0,t) &=& 0 %\hat k(1,1)\left[u_x(0,t) - \hat u_x(0,t)\right]
 \\  \label{eq-obsPDEBC1khat}
	\hat u(1, t) &=& 
 U(t) ,
\end{eqnarray}
with all NO gain kernels $\hat k = \hat{\mathcal K}(\lambda)$ of 
 %arbitrary desired 
   approximation accuracy $\epsilon\in (0,\epsilon^*)$ in relation to the exact backstepping kernel $k={\mathcal K}(\lambda)$ ensure that 
 %there exist $M, c^*>0$ such that 
 the observer error system, for all $u_0, \hat u_0 \in L^2[0,1]$, satisfies the exponential stability bound 
 %\eqref{eq-expstabbound}.
%Thus we arrive at
\begin{equation}\label{eq-expstabboundobs}
\|u(t)-\hat u(t)\| \leq M {\rm e}^{-(t-t_0)/2} \|u_0-\hat u_0\|, \quad\forall t\geq t_0,
\end{equation}
where $M(\epsilon,\bar\lambda)$ is defined in \eqref{eq-Mepslam}. 
\end{thm}
 
\begin{pf}
We start by postulating a PDE backstepping observer in the form
\begin{eqnarray}
    \label{eq-obsPDE}
	\hat u_t(x,t) &=& \hat u_{xx}(x,t) + \lambda(x) \hat u(x, t) 
 \nonumber\\
 &&+p_1(x)\left[u_x(1,t) - \hat u_x(1,t)\right]
 %, \qquad \mbox{$x\in[0,1)$}
 \\  \label{eq-obsPDEBC0}
 \hat u(0,t) &=& 0 % p_0\left[u_x(1,t) - \hat u_x(1,t)\right]
 \\  \label{eq-obsPDEBC1}
	\hat u(1, t) &=& 
 U(t). 
\end{eqnarray}
The observer error $\tilde u(x,t) = u(x,t) - \hat u(x,t)$ is governed by the system
\begin{eqnarray}
    \label{eq-erobsPDE}
	\tilde u_t(x,t) &=& \tilde u_{xx}(x,t) + \lambda(x) \tilde u(x, t) 
 \nonumber\\
 &&-p_1(x)\tilde u_x(1,t) 
 %, \qquad \mbox{$x\in[0,1)$}
 \\  \label{eq-erobsPDEBC0}
 \tilde u(0,t) &=& 0 % -p_0\tilde u_x(1,t) 
 \\  \label{eq-erobsPDEBC1}
	\tilde u(1, t) &=& 0. 
\end{eqnarray}
The backstepping transformation 
\begin{equation}
\label{eq-obsbkst}
    \tilde u(x,t) = \tilde w(x,t) - \int_x^1 p(x,y) \tilde w(y,t) dy
\end{equation}
converts \eqref{eq-erobsPDE}, \eqref{eq-erobsPDEBC0}, \eqref{eq-erobsPDEBC1} into
\begin{eqnarray}
    \label{eq-werobsPDE}
	\tilde w_t(x,t) &=& \tilde w_{xx}(x,t) 
 \\  \label{eq-werobsPDEBC0}
 \tilde w(0,t) &=& 0 
 \\  \label{eq-werobsPDEBC1}
	\tilde w(1, t) &=& 0 
\end{eqnarray}
provided $p(x,y)$ satisfies
\begin{equation} \label{eq-kyx}
    p(x,y) = k(y,x)
\end{equation}
with $k$ that is governed by \eqref{eq-kPDE}, \eqref{eq-kPDEBC0}, \eqref{eq-kPDEBCx}, and with the observer gain
\begin{equation}
    p_1(x) = -k(1, x). %, \quad p_0=k(1,1).
\end{equation}
It is crucial to note in \eqref{eq-kyx} that the arguments $x$ and $y$ have been commuted in $k(\cdot,\cdot)$. The commuting of the spatial arguments of the backstepping kernel is akin to transposing matrices in going between designs for controllers and observers in finite-dimensional LTI systems. The commuted order of the arguments $x$ and $y$ continues in the rest of the proof. The observer \eqref{eq-obsPDE}, \eqref{eq-obsPDEBC0}, \eqref{eq-obsPDEBC1} is next rewritten as
\begin{eqnarray}
    \label{eq-obsPDEk}
	\hat u_t(x,t) &=& \hat u_{xx}(x,t) + \lambda(x) \hat u(x, t) 
 \nonumber\\
 &&-k(1, x)\left[u_x(1,t) - \hat u_x(1,t)\right]
 %, \qquad \mbox{$x\in[0,1)$}
 \\  \label{eq-obsPDEBC0k}
 \hat u(0,t) &=& 0 %k(1,1)\left[u_x(0,t) - \hat u_x(0,t)\right]
 \\  \label{eq-obsPDEBC1k}
	\hat u(1, t) &=& 
 U(t) 
\end{eqnarray}
and the transformation \eqref{eq-obsbkst} as
\begin{equation}
\label{eq-obsbkstkhat}
    \tilde u(x,t) = \tilde w(x,t) - \int_x^1 k(y,x) \tilde w(y,t) dy.
\end{equation}
Theorem \ref{cor-DeepONet} applies to the kernel $k(y,x)$ of the observer backstepping transformation and the observer gains $-k(1,x)$. The observer 
\eqref{eq-obsPDEk}, \eqref{eq-obsPDEBC0k}, \eqref{eq-obsPDEBC1k} is henceforth implemented with the approximate kernel $\hat k$ as in \eqref{eq-obsPDEkhat}, \eqref{eq-obsPDEBC0khat}, \eqref{eq-obsPDEBC1khat}
whereas the backstepping transformation \eqref{eq-obsbkstkhat} is applied with $\hat k$ as
\begin{equation}
\label{eq-obsbkstkhathat}
    \tilde u(x,t) = \omega(x,t) - \int_x^1 \hat k(y,x) \omega(y,t) dy.
\end{equation}
The target system under the approximate kernel $\hat k$ becomes
\begin{eqnarray}\label{eq-target-omega}
    \omega_t(x,t) &=& \omega_{xx}(x,t) 
    +\Omega_0(x,t)+\Omega_1(x,t)
    % -\delta_{k0}(x) \omega(x,t)
    % \nonumber\\
    % &&- \int_x^1 \delta_{k1} (y,x) \omega(y,t) dy
    \\
    \omega(0,t) &=&0
       \\
    \omega(1,t) &=&0,
\end{eqnarray}    
where 
\begin{eqnarray}
    \Omega_0(x,t) &=& \delta_{k0}(x) \omega(x,t) + \int_x^1 \hat l(y,x)\delta_{k0} (y) \omega(y,t) dy
    \\
    \Omega_1(x,t) &=&\int_x^1 \left(\delta_{k1} (y,x) \omega(y,t) \right.
    \nonumber\\
    && \left. + \hat l(y,x)\int_y^1 \delta_{k1} (s,y) \omega(s,t) ds \right) dy
\end{eqnarray}
and 
$\delta_{k0}, \delta_{k1}$ are defined in \eqref{eq-deltakxy0}, \eqref{eq-deltakxy1}, with bounds  \eqref{eq-bounddeltak0}, \eqref{eq-bounddeltak1}. Note that the arguments in $\delta_{k1}$ have been commuted in the integral in \eqref{eq-target-omega}. 
%Recall that the bounds \eqref{eq-bounddeltak0}, \eqref{eq-bounddeltak1} hold on the perturbation kernels $\delta_{k0}, \delta_{k1}$. 
Similar as in the proof of Theorem \ref{thm-stabDeepONet-gf}, the Lyapunov functional 
\begin{equation}\label{eq-Vomega}
V = {1\over 2} \|\omega\|^2
\end{equation}
has a derivative 
\begin{equation}
\dot V \leq -{1\over 4} V, 
\end{equation}
namely, $V(t)\leq V_0 {\rm e}^{-(t-t_0)/4}$, provided $\epsilon \in [0,\epsilon^*]$, with $\epsilon^*$ obtained from \eqref{eq-delta*}. 
% \begin{equation}
% \dot V \leq -{1\over 2} (1 - \|\delta_{k0}\|_\infty-\|\delta_{k1}\|_\infty) V
% \leq -{1\over 2} (1 - 2\epsilon %(1+\bar \lambda)
% ) V
% \end{equation}
% Hence, for all $\epsilon \in \left[0,{1\over 
% 4 %2(1+\bar\lambda)
% }\right]$,
% $\dot V \leq -{1\over 4} V$, and, therefore, $V(t)\leq V_0 {\rm e}^{-(t-t_0)/4}$. 
The result \eqref{eq-obsPDEkhat} follows from \eqref{eq-Vomega}, \eqref{eq-obsbkstkhathat}, \eqref{eq-khatbound}, 
and the inverse backstepping transformation
\begin{equation}
\label{eq-obsbkstkhathat-inv}
    \omega(x,t) =  \tilde u(x,t)+ \int_x^1 \hat l(y,x) \tilde u(y,t) dy 
\end{equation}
whose kernel $\hat l$ satisfies the bound \eqref{eq-lhatbound}. 
%\eqref{eq-obsbkstkhathat-inv}, 
%\eqref{eq-lhatbound}. 
\qed
\end{pf}

\section{Collocated Output-Feedback Stabilization}
\label{sec-bkst-output-feedback}

In this section we put together the observer \eqref{eq-obsPDEkhat}, \eqref{eq-obsPDEBC0khat}, \eqref{eq-obsPDEBC1khat}, 
along with the observer-based controller
\begin{equation}\label{eq-obsfbk}
U(t) = \int_0^1 \hat k(1,x) \hat u(x,t) dx
\end{equation}
to stabilize the system \eqref{eq-PDE}, \eqref{eq-PDEBC0}, \eqref{eq-PDEBC1}  by output feedback.

The backstepping transformations %\eqref{eq-obsbkstkhathat} and 
\begin{eqnarray}\label{eq-outfbkotransobs}
    {\check w}(x,t) &=& \hat u(x,t) - \int_0^x \hat k(x,y) \hat u(y,t) dy\\
    \label{eq-outfbkotransobserr}
        \tilde u(x,t) &=& \omega(x,t) - \int_x^1 \hat k(y,x) \omega(y,t) dy.
\end{eqnarray}
%to 
transform the overall system into the cascade
\begin{eqnarray}
%\label{eq-target-wcheck}
\label{eq-wcheck1}
    \check w_t(x,t) &=& \check w_{xx}(x,t) 
%    \nonumber\\ &&
    + \delta_{k0} (x) 
    %\hat u(y,t) 
    \check w(x,t) dy
     \nonumber\\
    &&
    + \delta_{k0} (x,y) \int_0^x \hat l(x,y) \check w(y,t) dy \nonumber\\
    &&+ \int_0^x \delta_{k1} (x,y) 
    %\hat u(y,t) 
    \check w(y,t) dy
     \nonumber\\
    &&
    + \int_0^x \delta_{k1} (x,y) \int_0^y \hat l(y,\eta) \check w(\eta,t) d\eta dy
    \nonumber \\
    && -  \left(\hat k(1,x)-\int_0^x \hat k(x,y) \hat k(1, y) dy \right) \omega_x(1,t) %\tilde u_x(1,t)
    \nonumber\\ 
    \\
\label{eq-wcheck2}
    \check w(0,t) &=&0
       \\
\label{eq-wcheck3}
    \check w(1,t) &=&0
    \\  \label{eq-om1}
        \omega_t(x,t) &=& \omega_{xx}(x,t) 
        +\delta_{k0}(x) \omega(x,t) 
        \nonumber\\
        &&+ \int_x^1 \hat l(y,x)\delta_{k0} (y) \omega(y,t) dy
    \nonumber \\
    &&+\int_x^1 \left(\delta_{k1} (y,x) \omega(y,t) \right.
    \nonumber\\
    && \left. + \hat l(y,x)\int_y^1 \delta_{k1} (s,y) \omega(s,t) ds \right) dy
        %\int_x^1 \delta_k (y,x) \omega(y,t) dy
    \\  \label{eq-om2}
    \omega(0,t) &=&0
       \\  \label{eq-om3}
    \omega(1,t) &=&0.
\end{eqnarray}    
Both the $\omega$-subsystem \eqref{eq-om1}--\eqref{eq-om3}, which is autonomous, and the $\check w$-subsystem \eqref{eq-wcheck1}--\eqref{eq-wcheck3}, which is driven by the output $\omega_x(1,t)$ of the $\omega$-subsystem, are exponentially stable in $L^2[0,1]$ and higher norms for sufficiently small $\epsilon$. However, because the trace term $\omega_x(1,t)$ in the last line of \eqref{eq-wcheck1} cannot be easily bounded even by an $H^2$ norm of $\omega$, we do not pursue a  stability analysis of the composite system, i.e., we leave the ``separation principle'' unproven for the  
observer-based feedback  \eqref{eq-obsPDEkhat}, \eqref{eq-obsPDEBC0khat}, \eqref{eq-obsPDEBC1khat}, \eqref{eq-obsfbk}
acting on the system  \eqref{eq-PDE}, \eqref{eq-PDEBC0}, \eqref{eq-PDEBC1}. 
The technical challenge has nothing to do with the NO implementation of the kernel $\hat k$, as the challenge does not arise due to the perturbation kernels $\delta_{k0}, \delta_{k1}$. The challenge is due to the unbounded nature of the {\em output mapping} $\omega(t) \mapsto\omega_x(1,t)$, a challenge not encountered in ODEs but only in PDE control with boundary sensing or actuation. 
%In fact, an NO-based output-feedback design would be provably exponentially stabilizing in the $H^1$ norm $\|u(t)\|_{H^1} + \|\hat u(t)\|_{H^1}$ for the setups in rows three, four, seven, and eight, and in the $H^2$ norm $\|u(t)\|_{H^2} + \|\hat u(t)\|_{H^2}$ for the setups in rows five and six, in both cases with the aid of Agmon's inequality in the Lypunov analysis. However, those designs are more complicated, and the associated analysis for the full-state feedback and observer designs is also more complicated. And technical complications are contrary to our pedagogical objectives in this introductory exposition to NO-based PDE backstepping. Hence, we opt for the simplest setup, sacrificing the separation principle to gain clarity. 

The result given next, which is of a slightly more complicated form, is provable but we give it without a proof because the calculations are very, very lengthy and partly duplicate the calculations in the previous sections. The actuation-sensing setup is from the last row of Table \ref{table-contributions}, namely, a collocated Neumann actuation and Dirichlet sensing. Stability established is in the $H^1$ norm $\|u(t)\|_{H^1} + \|\hat u(t)\|_{H^1}$. 

\begin{thm}
\label{thm-stabDeepONet-gf-obsfbk}
Consider the system
\begin{eqnarray}
    \label{eq-PDEN}
	u_t(x,t) &=& u_{xx}(x,t) + \lambda(x) u(x, t) , \qquad \mbox{$x\in[0,1)$}
 \\  \label{eq-PDEBC0N}
 u(0,t) &=& 0
 \\  \label{eq-PDEBC1N}
	u_x(1, t) &=& U(t) 
\end{eqnarray}
with a measured Dirichlet output $u(1,t)$, along with the collocated observer-based Neumann-actuated controller
 \begin{eqnarray}
    \label{eq-obsPDEkhatN}
	\hat u_t(x,t) &=& \hat u_{xx}(x,t) + \lambda(x) \hat u(x, t) 
 \nonumber\\
 &&+ 
 %\left. \hat k_\xi(\xi, x)\right|_{\xi=1}
 \kappa(x)
 \left[u(1,t) - \hat u(1,t)\right]
 %, \qquad \mbox{$x\in[0,1)$}
 \\  \label{eq-obsPDEBC0khatN}
 \hat u(0,t) &=& 0 %\hat k(1,1)\left[u_x(0,t) - \hat u_x(0,t)\right]
 \\  \label{eq-obsPDEBC1khatN}
	\hat u_x(1, t) &=& U(t) -\hat k(1,1) \left(u(1,t) - \hat u(1,t)\right)\\
 U(t) &=& \hat k(1,1) u(1,t) + \int_0^1 
 \kappa(x)
 %\left. \hat k_\xi(\xi,x)\right|_{\xi=1} 
 \hat u(x,t) dx, 
\end{eqnarray}
where the gain function of both the controller and the observer is given by
\begin{equation}
    \kappa(x) := \left. \hat k_\xi(\xi,x)\right|_{\xi=1}. 
\end{equation}
For all $B_\lambda, B_{\lambda'}>0$ and for all $\lambda\in C^1([0,1])$ whose derivative is Lipschitz and which satisfies
$\|\lambda\|_\infty\leq B_\lambda$ and $ \|\lambda'\|_\infty\leq B_{\lambda'}$, there exists a  sufficiently small $\epsilon^*(B_\lambda, B_{\lambda'})>0$ such that for all NO gain kernels $\hat k = \hat{\mathcal K}(\lambda)$ of 
 %arbitrary desired 
   approximation accuracy $\epsilon\in (0,\epsilon^*)$ in relation to the exact backstepping kernel $k={\mathcal K}(\lambda)$ there exists sufficiently large $M(\epsilon,\bar\lambda)>0$ such that the above observer-based feedback  
%\eqref{eq-obsPDEkhat}, \eqref{eq-obsPDEBC0khat}, \eqref{eq-obsPDEBC1khat}, \eqref{eq-obsfbk}
 ensures that 
 %there exist $M, c^*>0$ such that 
 the closed-loop system, for all $u_0, \hat u_0 \in H^1[0,1]$, satisfies the exponential stability bound 
 %\eqref{eq-expstabbound}.
%Thus we arrive at
\begin{equation}\label{eq-expstabboundH1}
\|u(t)\|_{H^1} + \|\hat u(t)\|_{H^1} \leq M {\rm e}^{-(t-t_0)/4} \left(\|u_0\|_{H^1} +\|\hat u_0\|_{H^1} \right)
\end{equation}
for all $ t\geq t_0$. 
\end{thm}

The proof is based on the backstepping transformations \eqref{eq-outfbkotransobs}, \eqref{eq-outfbkotransobserr} into a perturbed version of the target system 
\begin{eqnarray}
%\label{eq-target-wcheckN}
\label{eq-wcheck1N}
    \check w_t(x,t) &=& \check w_{xx}(x,t) 
    \nonumber\\ &&
+ \left(\kappa(x)-\int_0^x \kappa(y)\hat k(x,y)  dy \right) \omega(1,t) %\tilde u_x(1,t)
    \\
\label{eq-wcheck2N}
    \check w(0,t) &=&0
       \\
\label{eq-wcheck3N}
    \check w(1,t) &=&0
    \\  \label{eq-om1N}
        \omega_t(x,t) &=& \omega_{xx}(x,t) 
    \\  \label{eq-om2N}
    \omega(0,t) &=&0
       \\  \label{eq-om3N}
    \omega(1,t) &=&0. 
\end{eqnarray}
The perturbation terms are as in \eqref{eq-wcheck1} and \eqref{eq-om1}, employing the functions $\delta_{k0}$ and $\delta_{k1}$ (which are uniformly bounded by $\epsilon$). The Lyapunov analysis employs the $H^1$ norms of $\check w$ and $\omega$, along with Agmon's inequality to bound the perturbation term $\omega(1,t)$ in the $\check w$-system using the norm $\|\omega_x\|$. 

\begin{figure*} 
    \centering
    \includegraphics{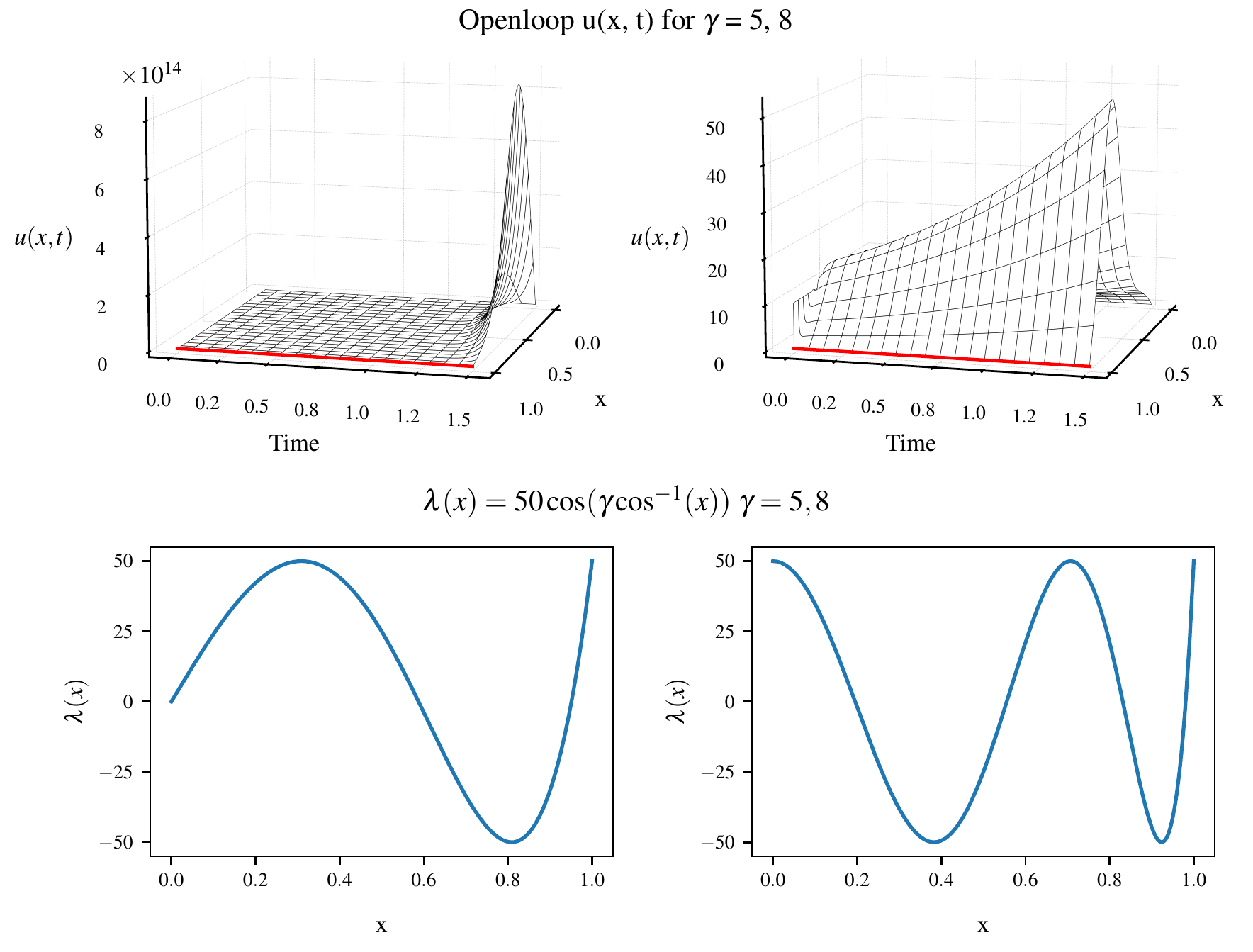}
    \caption{Open-loop instability (top) for the two respective reaction coefficients $\lambda(x)$ shown on the bottom row.}
    \label{fig:open-loop}
\end{figure*}

\begin{figure*} 
    \centering
    \includegraphics{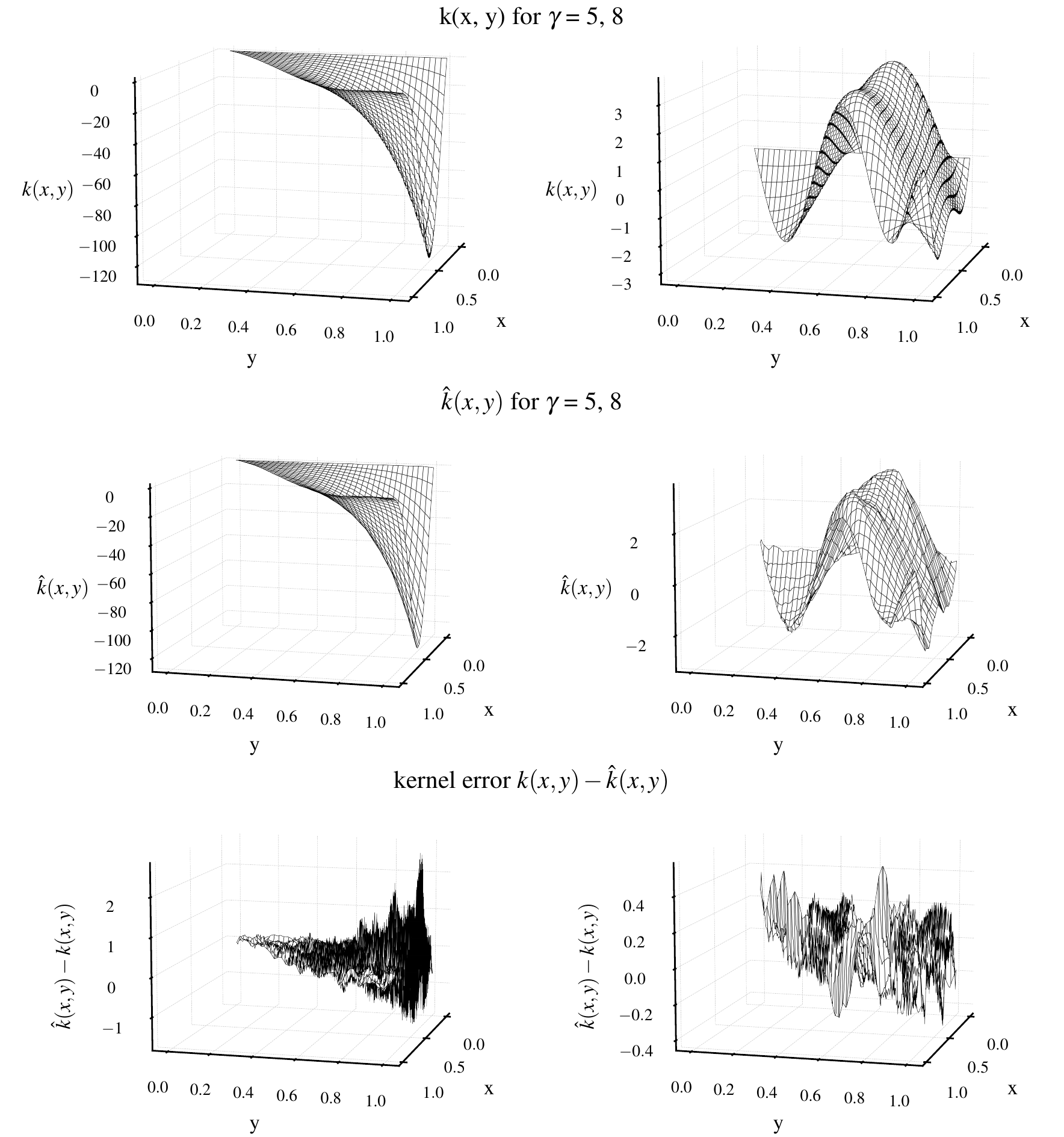}
    \caption{Examples of the kernel $k(x, y)$ (top row), learned kernel $\hat{k}(x, y)$ (middle row), and the kernel error $k(x, y) - \hat{k}(x, y)$ (bottom row). The two respective $\lambda (x)$ values correspond to the same respective values as in Fig \ref{fig:open-loop}.}
    \label{fig:kernels}
\end{figure*}

\begin{figure*} 
    \centering
    \includegraphics{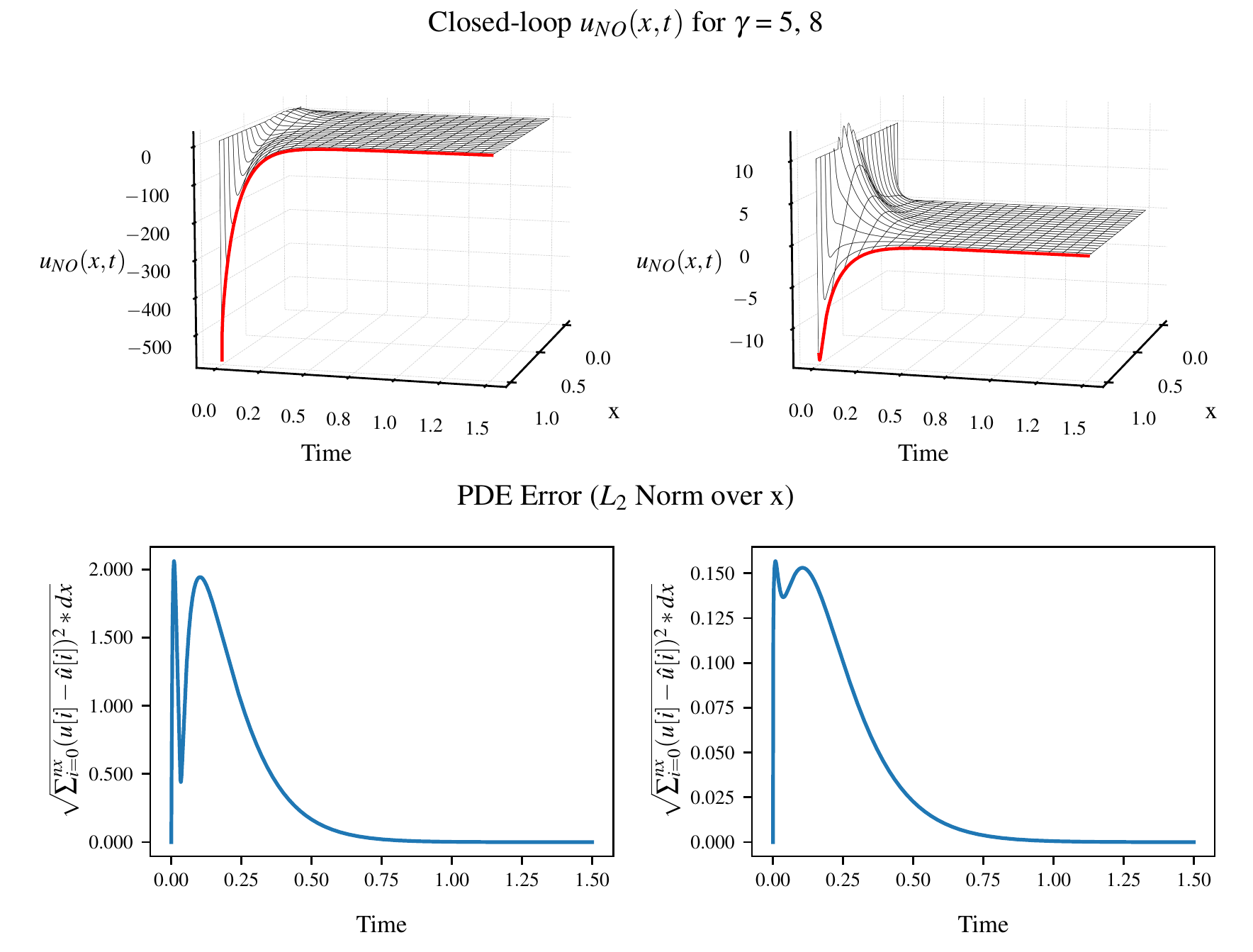}
    \caption{For the two respective $\lambda(x)$ values as in Fig \ref{fig:open-loop}, the top row showcases closed-loop solutions with the learned kernel $\hat{k}(x, y)$, whereas the bottom row shows the closed-loop PDE error between applying the original kernel $k(x, y)$ and the learned kernel $\hat{k}(x, y)$. }
    \label{fig:closed-loop}
\end{figure*}

\begin{figure*} 
    \centering
    \includegraphics{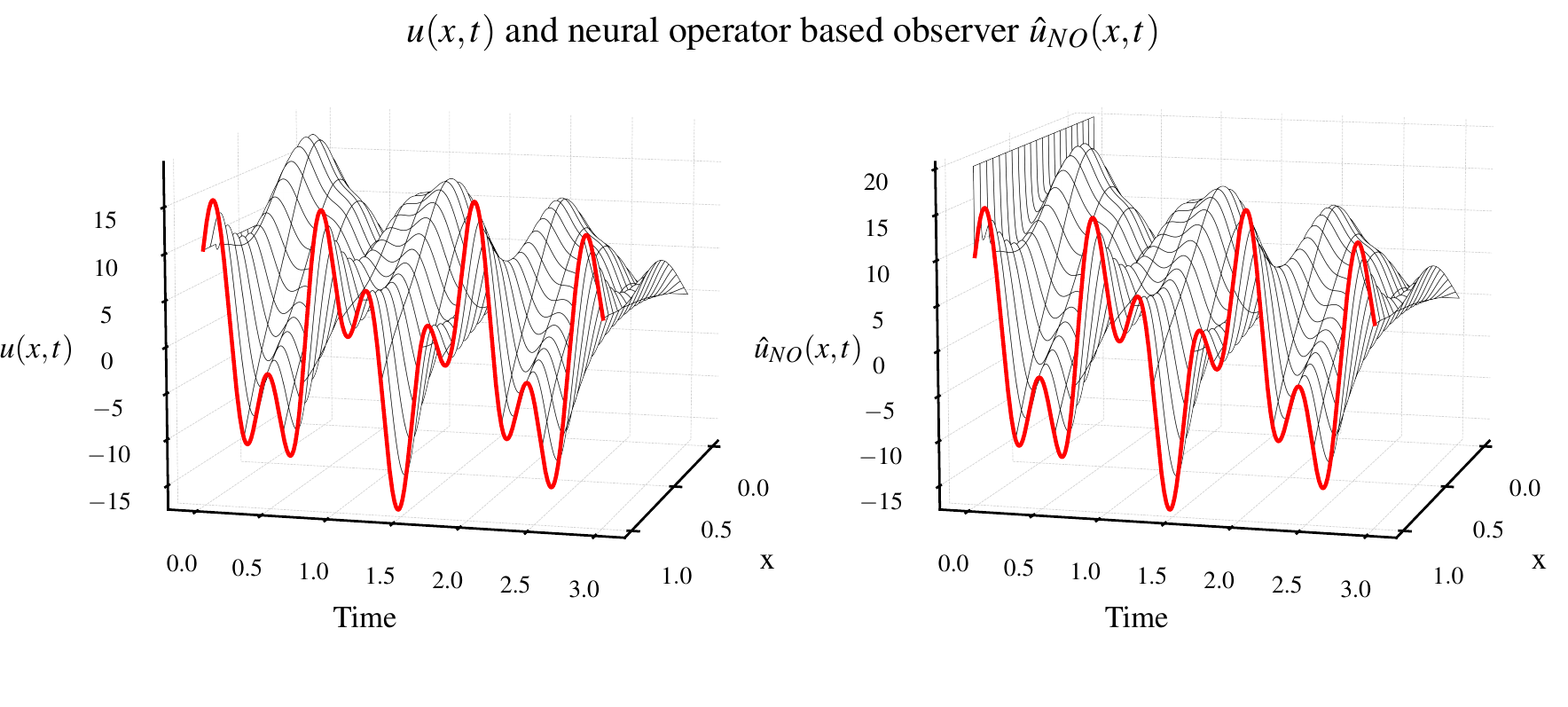}
    \caption{Left: PDE solution with $\lambda(x) = 20 \cos (5 \cos^{-1}(x))$ and  $U(t) = 7\sin(16\pi t) + 10 \cos(2\pi t)$. Right: the observer with the neural operator-learned kernel. Note the difference between the plant initial condition $u(x, 0) = 10$ and the twice as large initial condition of the exact and neural operator observers, $\hat{u}_{\rm NO}(x, 0) = \hat{u}(0, x) =20$. The peak error between the analytical observer (not shown) and the neural operator observer is around 0.3. }
    \label{fig:observer}
\end{figure*}

\section{Numerical Results: Full-State Feedback, Observer, and Output Feedback}
\label{sec-numerical}
In Figure \ref{fig:open-loop}, we show that the system is open-loop unstable for the reaction term $\lambda(x) = 50\cos(\gamma \cos^{-1}(x))$ for $\gamma=5, 8$. The increased oscillation in larger $\gamma$ yields a lower rate of instability as shown on the right. We simulate the PDE and its control using the finite difference scheme in Appendix \ref{appendix-fd}. 

In Figure \ref{fig:kernels}, we demonstrate both the analytical and learned DeepONet kernels for the two $\gamma$ values corresponding to Figure \ref{fig:open-loop}. To learn the mapping $\mathcal{K}: \lambda(x) \mapsto k(x, y)$, we construct a dataset of 900 different $\lambda(x)$ as the Chebyshev polynomials define above with $\gamma \sim $ uniform $(4, 9)$. We choose $\lambda$ of this form due to the rich
set of kernel functions generated by varying only a single parameter. To effectively utilize the DeepONet without modifying the grid, we stack $\lambda(x)$ repeatedly $n_y$ times over the $y$ axis to make a 2D input to the network. Then, we capitalize on the 2D mapping by implementing a CNN for the DeepONet branch network. In the future, one can explore neural operators on irregular girds along the direction of \cite{https://doi.org/10.48550/arxiv.2207.05209}. For training, the relative $L_2$ error is $3.5e-2$ and the testing error is $3.6e-2$. With the learned neural operator, we achieve speedups on the magnitude of $10^3$ compared to an efficient finite difference implementation. In Figure \ref{fig:closed-loop}, we demonstrate closed-loop stability with the neural operator approximated gain function for the control feedback law. Additionally, we see the error is largest at the beginning achieving a maximum in both cases of approximately $10\%$. 

In Figure \ref{fig:observer}, we test the observer \eqref{eq-obsPDEBC0khat}, \eqref{eq-obsPDEBC0khat}, \eqref{eq-obsPDEBC1khat} with a DeepONet-approximated kernel trained as above using $\lambda(x) = 20 \cos(5 \cos^{-1}(x))$ with $\gamma \sim $ uniform $(4, 9)$. Additionally, we apply a boundary signal of $U(t) = 7 \sin(16\pi t ) + 10 \cos(2\pi t)$ to generate a challenging and rich PDE motion for estimation. The true system state begins with initial conditions $u(x, 0) = 10$ while the DeepONet observer has initial conditions of $\hat{u}_{NO}(x, 0) = 20$. Despite this, the observer approximates the PDE well with a peak error of less than $5\%$ compared to the analytical observer while maintaining the same $10^3$x speedup over the finite difference scheme.

\section{Conclusions}
\label{sec-conclusions}
In this paper, we build on the framework introduced in \cite{https://doi.org/10.48550/arxiv.2302.14265}, and depicted in Figures \ref{fig:0} and \ref{fig:deeponetdiagram}, and extend the neural operator-supported PDE backstepping methodology from the hyperbolic to the harder parabolic case. We limit ourselves to the reaction-diffusion parabolic class for the clarity of exposition. 

With the foundation laid for hyperbolic and parabolic PDE backstepping designs, which free the user from having to solve kernel PDEs in real time and result in a {\em thousandfold speedup}, the road is open to developing this methodology for two important control domains in which the backstepping kernels constantly evolve in the course of implementation: (1) gain scheduling for nonlinear PDEs, where the kernel depends on the current state of the PDE; and (2) adaptive control of PDEs whose functional coefficients are unknown, have to be adaptively estimated online, and the kernel has to be continuously updated. In both applications, the solving of the $k$-PDE online is eliminated with the aid of the pre-determined neural operator $\hat{\mathcal{K}}$. 

\appendix
\section*{Appendix}
\section{Neural networks notation}\label{sec-NN}

An $n$-layer neural network (NN) $f^\mathcal{N}:\mathbb{R}^{d_1}\rightarrow \mathbb{R}^{d_n}$ is given by
%of $n$ layers is a mapping from $\mathbb{R}^{d_1}$ to $\mathbb{R}^{d_n}$ with the following structure
\begin{eqnarray} \label{eq-nn}
    f^\mathcal{N}(x, \theta) := (l_n \circ l_{n-1} \circ ... \circ l_2 \circ l_1) (x,\theta)
\end{eqnarray}
where layers $l_i$ start with $l_0 = x\in\mathbb{R}^{d_1}$ and continue as 
%is of the form
\begin{equation}
%    l_1(x,\theta_1,\ldots,\theta_{i+1}) &:=& \sigma(W_1 x + b_1) \\ 
    l_{i+1}%(x,\theta_1,\ldots,\theta_{i+1})
    (l_i,\theta_{i+1}):= \sigma(W_{i+1} l_i + b_{i+1}), \quad i=1,\ldots,n-1
    % \\
    % l_n &:=& \sigma(W_n l_{n-1} + b_n) 
\end{equation}
%where $l_0 = x\in\mathbb{R}^{d_1}$, 
$\sigma$ is a nonlinear activation function, and weights $W_{i+1} \in \mathbb{R}^{d_{i+1} \times d_i}$ and biases $b_{i+1} \in \mathbb{R}^{d_{i+1}}$ are parameters to be learned, collected into $\theta_i\in \mathbb{R}^{d_{i+1}(d_i+1)}$, 
and then 
%. All of $W_1 \in \mathbb{R}^{d_{2} \times d_1},\ldots, W_n \in \mathbb{R}^{d_n \times d_{n-1}}, b_1 \in \mathbb{R}^{d_1}, \ldots, b_n \in \mathbb{R}^{d_n}$ are collected 
into $\theta = [\theta_1^{\rm T},\ldots, \theta_n^{\rm T}]^{\rm T}\in\mathbb{R}^{\sum_{i=1}^{n-1} d_{i+1}(d_i+1)}$.
%We enforce that the first weight matrix is of the form $W_1 \in \mathbb{R}^{d_{2} \times d_1}$ and the final weight matrix and bias term are $W_n \in \mathbb{R}^{d_n \times d_{n-1}}$ and $b_n \in \mathbb{R}^{d_n}$. 
Let $\vartheta^{(k)}, \theta^{(k)} \in\mathbb{R}^{\sum_{i=1}^{k-1} d_{k,(i+1)}(d_{k,i}+1)}$ denote a sequence of NN weights.

A neural operator (NO) 
for approximating a nonlinear operator $\mathcal{G}: {\mathcal U} \mapsto {\mathcal V}$ is defined as 
\begin{eqnarray}
    {\mathcal G}_\mathbb{N}(\mathbf{u}_m)(y) = \sum_{k=1}^p g^\mathcal{N} (\mathbf{u}_m; \vartheta^{(k)}) f^\mathcal{N} (y; \theta^{(k)})
\end{eqnarray}
where ${\mathcal U}, {\mathcal V}$ are function spaces of continuous functions $u \in {\mathcal U}, v \in {\mathcal V}$.
$\mathbf{u}_m$ is the evaluation of function $u$ at points $x_i=x_1, ..., x_m$, $p$ is the number of chosen basis components in the target space, $y \in Y$ is the location of the output function $v(y)$ evaluations, %in the domain of the target function space $v(y) \in \mathcal V$, 
and $g^\mathcal{N}$, $f^\mathcal{N}$ are NNs termed  branch and trunk networks. Note, $g^\mathcal{N}$ and $f^\mathcal{N}$ are not limited to feedforward NNs \eqref{eq-nn}, but can also be of %many forms such as 
convolutional or recurrent.
%depending on data structure. 

% where $\hat u(x,t) = \check w(x,t) + \int_0^x \hat l(x,y) \check w(y,t) dy$, and then either by direct Lyapunov analysis, as in \cite{SMYSHLYAEV2005613,Smyshlyaev2010}, or by input-to-state stability analysis \cite{Karafyllis-Krstic-book-ISS-Springer}, in both cases in $H^1$, and for sufficiently small $\epsilon$. \qed
% \end{pf}

 % \cite{1369395}
% Closed-form boundary State feedbacks for a class of 1-D partial integro-differential equations

% \cite{krstic2008boundary}
% Boundary Control of {PDE}s: A Course on Backstepping Designs

% \cite{Smyshlyaev2010}
% Adaptive Control of Parabolic {PDE}s

\section{FD Scheme for Goursat-Form Kernel PDE} \label{appendix-fd}
For the PDE in \eqref{eq-PDE}, \eqref{eq-PDEBC0}, \eqref{eq-PDEBC1}, we utilize the following finite difference scheme adapted from \cite{Smyshlyaev2010}:
\begin{eqnarray}
    k_j^{i+1} &=& -k_j^{i-1} + k_{j+1}^i + k_{j-1}^i + h^2\lambda_j\frac{k_{j+1}^i + k_{j-1}^i}{2} \\ 
    k^{i+1}_i &=& k_i^i + \frac{h}{2} \lambda_i \\ 
    k_{i+1}^{i+1} &=& k_i^i - \frac{h}{4} (\lambda_i + \lambda_{i+1}), \qquad k_1^{j+1} = 0
\end{eqnarray}
with $k_i^j = k((i-1)h, (j-1)h), i=2, ..., N, j=2, ..., i-1, \lambda_i=\bar{\lambda}((i-1)h), h=1/N$ where $N$ is the number of spatial steps.

\bibliography{refs}
%\bibstyle{abbrv}
\bibliographystyle{abbrv}

\end{document}